\documentclass{aastex}
\usepackage{amsmath,amssymb,bm}
\begin{document}

\title{The Kozai--Lidov Mechanism in Hydrodynamical Disks. II. Effects of binary and disk parameters}
\author{Wen Fu{$^{1,2}$}, Stephen H. Lubow{$^3$} and Rebecca G. Martin{$^4$}}
\affil{{$^1$}Department of Physics and Astronomy, Rice University, Houston, TX 77005, USA; wf5@rice.edu\\
{$^2$}Los Alamos National Laboratory, Los Alamos, NM 87545, USA\\
{$^3$}Space Telescope Science Institute, Baltimore, MD 21218, USA\\
{$^4$}Department of Physics and Astronomy, University of Nevada, Las Vegas, Las Vegas, NV 89154, USA}


\begin{abstract}
\cite{Martin14b} showed that a substantially misaligned accretion disk around one component of a binary system can undergo global damped Kozai--Lidov (KL) oscillations. During these  oscillations, the inclination and eccentricity of the disk are periodically exchanged.
However, the robustness of this mechanism and its dependence on the system parameters were unexplored. 
In this paper, we use three-dimensional hydrodynamical simulations to analyze how various
binary and disk parameters affect the KL mechanism in hydrodynamical disks. 
The simulations include the effect of gas pressure and viscosity, but ignore the effects
of disk self-gravity.
We describe results for different numerical resolutions, binary mass ratios and orbital eccentricities, initial disk sizes, 
initial disk surface density profiles, disk sound speeds, and disk viscosities. 
We show that the KL mechanism can operate for a wide range of binary-disk parameters. 
We discuss the applications of our results to astrophysical disks in various accreting systems. 
\end{abstract}

\keywords{accretion, accretion disks -- binaries: general -- hydrodynamics -- planets and satellites: formation}

\section{Introduction}

Disks in binary systems may be misaligned with respect
to their binary orbit planes. Alignment is expected
in cases where the disk formation process itself
is coplanar, such as the case of mass-exchange binaries.
In that case, a coplanar gas stream originates from the inner Lagrange
point of the mass-losing star and results in the formation of a coplanar disk.
Coplanarity is expected if the binary forms from disk
fragmentation \citep{Bonnell94, Offner10}. Furthermore, coplanarity is the long term
expected outcome of a misaligned disk, due to the effects of tidal dissipation
in the disk \citep{Papaloizou95, Bate00, Lubow00, King13}.

On the other hand, disk misalignment is expected to persist for some time period in
cases where the binary forms from a merger process,
such as occurs with supermassive black hole (SMBH) binary systems   \citep[e.g.,][]{King06}. 
If a young binary star system accretes material after its
formation process, the material is likely to be misaligned with the binary orbit and so misaligned disks may form around young stars \citep{Bate10}. In addition, a coplanar disk
may become noncoplanar due to an instability, such as
radiation warping \citep{Pringle96, Wijers99, Ogilvie01b}. Be/X-ray binaries  may have transient disks that are misaligned with respect 
to the binary orbit \citep{Martin09,Martin11}. These disks may be produced
by the expulsion of equatorial material 
from a rapidly rotating Be star whose spin axis is misaligned
with the binary rotation axis \citep[e.g.,][]{Porter03}.

There is both direct and indirect observational evidence for 
disk noncoplanarity. For widely separated stars in a binary, greater than 40 AU, 
a misaligned disk may occur because the stellar equatorial inclinations, based on spins, are observationally inferred to be misaligned with respect to the binary orbital planes \citep{Hale94}. 
The young binary system HK Tau provides
direct evidence for noncoplanarity, since disks are observed around
both components with one disk edge-on and the other more face-on
\citep{Stapelfeldt98}. Recent ALMA observations by \cite{Jensen14}
suggest that its disks are mutually misaligned by more than $60^{\circ}$, although
the plane of the binary orbit is not known. Strong mutual disk misalignment ($\sim 72^{\circ}$) has lately also been detected for the two circumstellar disks in V2434 Ori, a binary system in Orion \citep{Williams14}.
Less direct evidence comes from the existence
of extra-solar planets whose orbits are tilted with respect to the spin 
axis of the central star \citep{Albrecht12, Winn14}.  If the planets reside in binary star systems, this evidence suggests
that these planets may have formed in disks that are misaligned with the binary orbital plane \citep[e.g.,][]{Bate10, Batygin12}.

Test particles whose orbits are sufficiently inclined with respect to the plane of a circular orbit binary can undergo Kozai-Lidov (KL) oscillations \citep{Kozai62, Lidov62}. In these oscillations, the particle's orbital inclination and eccentricity evolve in such a manner that inclination is exchanged for eccentricity. Due to this process,  a test particle orbit that is initially circular can achieve a high eccentricity when it evolves to smaller inclination. For example, an initially circular  orbit at an inclination of $60^{\circ}$ achieves an eccentricity of about 0.75. Further extensions of this theory show that even stronger effects can occur for eccentric orbit binaries \citep{Ford00, Lithwick11, Naoz11, Naoz13a, Naoz13b, Teyssandier13, Li14, Liu15}.

The KL effect for ballistic objects has been applied to a wide range of astronomical processes. These include inclinations of asteroids and irregular satellites \citep{Kozai62, Nesvorny03},  high orbital eccentricities of some extra-solar planets and formation of hot Jupiters \citep{Holman97, Mazeh97, Wu03, Takeda05, Xiang13,  Dawson14, Dong14, Petrovich15, Rice15}, misalignment between an exoplanet's orbital axis and stellar spin axis \citep{Storch14, Storch15}, black hole mergers \citep{Blaes02, Miller02},  the formation of close binary stars \citep{Harrington68, Mazeh79, Kiseleva98, Fabrycky07}, tidal disruption events \citep{Chen11}, and the formation of Type Ia supernovae \citep{Kushnir13}.

In a recent {\it Letter} \citep[hereafter Paper I]{Martin14b}, we showed in our Smoothed Particle Hydrodynamics (SPH) simulations 
that a fluid disk with pressure and viscosity that orbits a member of a binary
can also undergo global KL oscillations.
In this paper, we explore this process in more detail by considering how it is affected
by various disk and binary parameters. Following the approach of Paper I,
we attempt to relate the properties of disk KL oscillations to the properties
of KL particle oscillations.

The outline of the paper is as follows. In Section \ref{sec:part}
we describe the properties of particle orbits in binaries.
The parameters of the particle orbits and the binaries are chosen to be similar to those of the disk simulations.
In Section \ref{sec:kld} we describe the results of SPH simulations
for a range of initial particle numbers, disk viscosities,
disk aspect ratios, initial density profiles (including changes to 
initial inner and outer disk radii),  initial disk inclinations, binary
mass ratios, and binary eccentricities.
Section \ref{sec:sum} contains the discussion and summary.


\section{KL cycle of a test particle \label{sec:part}}
As discussed in Paper I, the properties of a fluid disk undergoing KL oscillations are related
to the properties of test (massless) particles undergoing KL oscillations that orbit at similar radii. 
In both the disk and test particle cases, the inclination and eccentricity are periodically exchanged. However, there are some
fundamental differences.
The period of a test particle oscillation varies with distance from the central object.
In the case of a disk
that is subject to pressure and viscous forces,
the oscillations are global (i.e., the period is the same over all radii)  and subject to damping by dissipation, unlike the test particle case. 
The global disk oscillation frequency is expected to involve an angular momentum weighted average
of the test particle oscillation frequencies (Equation (4) of Paper I).
Before considering the disk oscillations, we describe in this section the behavior
of test particle orbits. We apply 
binary parameters that are similar to those
we adopt for the disk simulations described in Section \ref{sec:kld}.
 
We consider ballistic particle orbits around one component of a binary system. 
Each particle orbit is initially circular, but substantially inclined 
with respect to the orbital plane of the binary. If the binary orbit is circular, then
over long timescales, the binary companion  effectively acts as a uniform circular ring.
Since this effective potential is time independent,
the energy of the particle about the central mass is conserved. Thus, the semi-major axis $a$ of the particle in a binary is nearly independent of time $t$, that is
\begin{equation}
a(t)  \approx {\rm constant},
\label{eq:adot}
\end{equation}
where the constant is determined by initial conditions.
Because the time-averaged perturbing potential is axisymmetric, 
the component of the  angular momentum of the particle that is perpendicular to the binary orbital plane is approximately conserved over long
timescales. From this conservation principle and Equation (\ref{eq:adot})
we have that
\begin{equation}
\sqrt{1-e^2(t)} \, \cos{\left(i(t) \right)} \approx {\rm constant},
\label{eq:e-i}
\end{equation}
where the constant is determined by initial conditions, $e$  is the orbital eccentricity of the particle, and $i$ is the inclination of the particle orbit with respect to the binary orbital plane.

We denote the masses of the binary component objects as $M_{\rm c}$ and $M_{\rm p}$, where $M_{\rm c}$ ($M_{\rm p}$)  is the mass of the central (perturbing) object to the particle orbit.
Analytic calculations have determined that the KL oscillation period for a particle subject to the effect of a binary in an eccentric orbit is approximately given by
\begin{equation}
\frac{\tau_{\rm KL}}{P_{\rm b}}\approx \frac{M_{\rm c}+M_{\rm p}}{M_{\rm p}}\frac{P_{\rm b}}{P}(1-e^2_{\rm b})^{3/2}\label{eq:klperiod}
\end{equation}
\citep{Holman97, Innanen97, Kiseleva98}, where $P=\sqrt{G M_{\rm c}/a^3}$ is the orbital period of the particle with orbital semimajor axis $a$,  $P_{\rm b}=2\pi/\Omega_{\rm b}$ is the orbital period of the binary,  
$e_{\rm b}$ is the eccentricity of the binary orbit, and
$\Omega_{\rm b}=\sqrt{G(M_{\rm c}+M_{\rm p})/a_{\rm b}^3}$ is the binary orbital frequency for binary semi-major axis $a_{\rm b}$. 

For the KL oscillations to occur, the initial inclination of the test particle orbit $i_0$ must satisfy the condition that $\cos^2{(i_0)} < \cos^2{(i_{\rm cr})}  = 3/5$,
where $i_{\rm cr}$ is the critical angle for KL oscillations.
This condition requires that $39^\circ \la i_0 \la 141^\circ$.
During the KL cycle, the  inclination $i$ of an initially circular prograde particle orbit ($i_0 < 90^\circ$) oscillates between the initial value $i_0$ and the
critical angle $i_{\rm cr} \simeq 39^{\circ}$, while the eccentricity oscillates between $e_{\rm min}=0$ and the maximum value of
\begin{equation}
e_{\rm max}=\sqrt{1-\frac{5}{3}{\rm cos}^2(i_0}),
\label{eq:maxe}
\end{equation}
that is achieved for $i=i_{\rm cr}$,
which follows from Equation (\ref{eq:e-i}) with the right-hand side constant value of 
$\cos{(i_0)}$
\citep[e.g.,][]{Innanen97}.

For an eccentric binary orbit,  the vertical component of a particle's angular
momentum is not conserved. Equation (\ref{eq:e-i}) breaks down over long timescales and the maximum eccentricity of the particle 
can approach unity \citep{Ford00, Lithwick11, Naoz11, Naoz13a, Naoz13b, Liu15}.
The orbit can flip its orientation from prograde to retrograde 
with respect to the binary. This flip occurs as the orbit passes through a radial state ($e=1$). 
In passing through a nearly radial orbit, the particle may collide with the central object.

In our numerical calculations, we compute the eccentricity of a test particle with coordinates $\bm{r}(t)$ by considering its
specific angular momentum $\bm{j}(t)$  relative to the disk central object at position $\bm{r}_{\rm c}(t)$,
\begin{equation}
\bm{j}(t) = ( \bm{r} -\bm{r}_{\rm c})\bm{\times}(\bm{\dot r}-\bm{\dot r}_{\rm c}),
\label{j}
\end{equation}
where the dot denotes differentiation in time.
The relative specific energy of the particle is
\begin{equation}
E(t) = \frac{1}{2}|\bm{\dot r}-\bm{\dot r}_{\rm c}|^2-\frac{G M_{\rm c}}{|\bm{r}-\bm{r_c}|}
\end{equation}
and thus its eccentricity is
\begin{equation}
e(t) =\sqrt{1+\frac{2 E |\bm{j}|^2}{(GM_{\rm c})^2}} .
\label{eq:e}
\end{equation}
The evolving inclination of the particle is
\begin{equation}
i(t) =\arccos \left( \frac{j_z}{|\bm{j}|}\right),
\label{eq:i}
\end{equation}
where $j_z$ is the component of the angular momentum in
Equation~(\ref{j}) that is perpendicular to the orbital plane of the binary.

\begin{figure}
\centering
\includegraphics[width=0.95\textwidth]{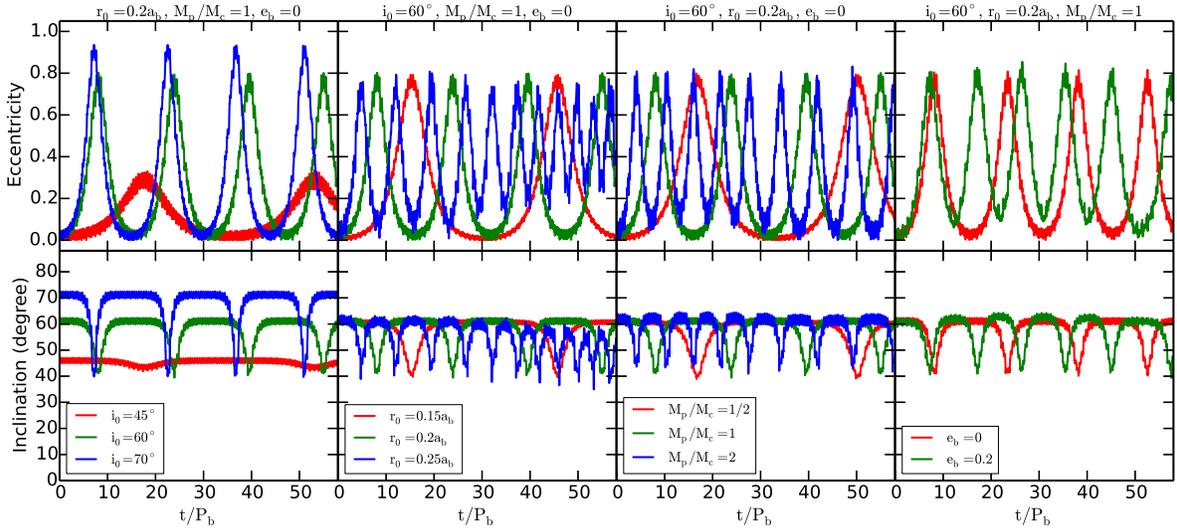}
\caption{Eccentricity (upper row) and inclination (bottom row) evolution of a test particle for different values of the initial orbital inclination $i_0$, initial orbital radius $r_0$, binary mass ratio $M_{\rm p}/M_{\rm c}$, and binary eccentricity $e_{\rm b}$. The binary orbit has a semi-major axis of $a_{\rm b}$. Time is in units of binary orbital period $P_{\rm b}$. These results are obtained from three-body numerical simulations. 
 (Color online) \label{fig:testpart}}
\end{figure}

Figure \ref{fig:testpart} plots the numerical results for a test particle undergoing KL oscillations from simple three-body simulations. 
The upper and lower rows show the evolution of the particle's orbital eccentricity and inclination, respectively. 
We investigate the effects of 
changes to four system parameters: the initial orbit inclination $i_0$, the initial orbital radius of the particle $r_0$, the binary mass ratio $M_{\rm p}/M_{\rm c}$, and
the binary eccentricity $e_{\rm b}$.  In all cases the initial orbital eccentricity of the particle is taken to be zero, $e_0=0$.
Each column of the figure corresponds to runs where we vary one of these parameters and fix the other three at fiducial (reference standard)
values. 
The fiducial parameters that we adopt for this figure 
are  $i_0 = 60^\circ, r_0 = 0.2 a_{\rm b}, e_{\rm b}=0,$ and  $M_{\rm p}/M_{\rm c}=1$.  
The fiducial value of $r_0$ is chosen as a characteristic radius 
that typically lies within the disks that we simulate  in Section 
\ref{sec:kld}.

The left hand panels show that when the orbit of the particle is highly inclined (well above critical KL angle 
$i_{\rm cr}$; see the green and blue lines), the KL cycle period 
is relatively insensitive to inclination, as expected by Equation (\ref{eq:klperiod}) that has no dependence on inclination.  The simulated period is sometimes close to that
predicted by Equation (\ref{eq:klperiod}). 
 For $i_0=70^\circ$ (blue line), the eccentricity oscillation has $e_{\rm max}\simeq 0.93$ and $\tau_{\rm KL}\simeq15 P_{\rm b}$, while Equations (\ref{eq:klperiod}) and (\ref{eq:maxe}) 
 predict that $e_{\rm max}=0.90$ and $\tau_{\rm KL}=16 P_{\rm b}$. 
 When the initial orbital tilt is just above the critical KL angle, the KL cycle period given by Equation (\ref{eq:klperiod}) becomes less accurate. 
 For $i_{0}=45^{\circ}$ (red line), the period from the simulation is $\tau_{\rm KL}\simeq 33 P_{\rm b}$, which is more than twice the analytical value. 
 
 The second column shows the effect of changing the initial orbital radius of the particle $r_0$. The KL periods for $r_0=0.15 a_{\rm b}$, $0.2 a_{\rm b}$, and $0.25 a_{\rm b}$ are $\sim 30 P_{\rm b}$,
  $\sim 15 P_{\rm b}$, 
 and $\sim 9 P_{\rm b}$, respectively. These values agree well (within $\sim 25\%$) with  the predictions of Equation (\ref{eq:klperiod}) that scales as $\tau_{\rm KL} \propto r^{-3/2}_0$ and produces KL periods of $24 P_{\rm b}$, $16 P_{\rm b},$ and $11 P_{\rm b}$, respectively.
 Note that with $r_0=0.25 a_{\rm b}$, the KL cycle period becomes shorter and the oscillation amplitude is reduced  at later time ($e_{\rm min}$ does not  reach zero). This is likely because at this orbital radius, the perturbing potential from the companion is more significant.
 As a result, the stronger influence makes the analytic description as a secular (gradual) effect less accurate and causes the oscillations to be irregular. 
 
 The effect of varying the binary mass ratio is shown in the third column.
  Equation (\ref{eq:klperiod}) predicts that $\tau_{\rm KL}/P_{\rm b} \propto (M_{\rm c}/M_{\rm p}+1)^{1/2}(M_{\rm c}/M_{\rm p})^{1/2}$ such that $\tau_{\rm KL}=10P_{\rm b}$, $16P_{\rm b}$, and $27P_{\rm b}$ for binary mass ratios $M_{\rm p}/M_{\rm c}=2$, $1$, and $1/2$, respectively. The results of the numerical simulations give us $\tau_{\rm KL}= 7P_{\rm b}$, $15P_{\rm b}$, and $32P_{\rm b}$. 
  The agreement is then within about 30\%.
  
 In the fourth column, we show the effect of varying the binary eccentricity. 
As discussed above, previous studies have shown that for eccentric orbit binaries,
 the maximum eccentricity can grow beyond $e_{\rm max}$ given by Equation~(\ref{eq:maxe}) and the maximum inclination
 can grow above the initial value of $i_0$.
As expected, for a somewhat higher binary eccentricity $e_{\rm b}=0.2$ (green lines), the maximum and minimum particle eccentricity sometimes increases somewhat
in each KL oscillation. 
The oscillation period is shorter than in the circular case, 
as predicted by Equation  (\ref{eq:klperiod}), since the companion object interacts more strongly with the particle at binary
periastron. According to Equation (\ref{eq:klperiod}), for $e_{\rm b}=0.2$, the KL oscillation period is approximately $14P_{\rm b}$, whereas the numerically determined period in Figure \ref{fig:testpart} is about $10P_{\rm b}$. Therefore, the reduction in the KL oscillation period due to binary orbital eccentricity is somewhat more severe than what is predicted analytically .


\section{KL cycle of a hydrodynamic disk \label{sec:kld}}

In this section, we describe the KL cycles of three-dimensional hydrodynamic disks. We use the SPH code  \texttt{PHANTOM} \citep{Lodato10, Price10, Nixon12a, Nixon12b,Price12,  Nixon13}.  The disk is initially circular, centered around the binary component of mass $M_{\rm c}$, and subject to perturbations by the other binary component 
of mass $M_{\rm p}$. The disk orbital plane is initially misaligned with respect to the binary orbital plane.  
The disk initially extends from radius $r_{\mathrm{in}}$ to radius
$r_{\mathrm{out}}$. The inner boundary of the simulated region is set to the disk initial inner disk radius 
$r_{\mathrm{in}}$. As particles move to $r \le r_{\mathrm{in}}$, they are removed from the simulation. 
We also impose an inner boundary radius around the perturbing companion, since some disk mass can be transferred to that component. The \texttt{PHANTOM} code adopts cubic spline kernel as the smoothing kernel. The softening length in the gravitational force is taken to be the same as the smoothing length. The number of neighbors is roughly constant at $N_{\rm neigh} \approx 58$. 

We define a certain set of fiducial (reference) model parameters. The disk is locally isothermal  with sound speed $c_{\rm s} \propto r^{-3/4}$ and disk aspect ratio $H/r = 0.035$ at the initial inner disk radius $r_{\mathrm{in}}$. 
These parameters allow both $\alpha$ and the smoothing length $\left<h\right> /H$ to be constant throughout the disk radius \citep{Lodato07}. 
We employ an explicit accretion disk viscosity which corresponds to an approximately constant  
$\alpha$ parameter \citep{SS73} over the disk \citep[see][]{Lodato10}.
The viscous stresses include a nonlinear term with a coefficient $\beta_{\rm AV} = 2$ (AV stands for artificial viscosity) that suppresses interparticle penetration, as is standard in SPH codes. 
The fiducial binary mass ratio is unity and the binary orbital eccentricity is zero (circular orbit).
This fiducial setup is similar to  that presented in Paper I. The only difference is that the initial number of particles in the fiducial model is set to $N=3\times10^5$, while Paper I
used $N=1 \times 10^6$. The lower value adopted here reduces computational overhead in carrying out the various simulations, while providing results that are well converged in $N$, as we discuss below.

\begin{deluxetable}{l c r r}
\tabletypesize{\footnotesize}
\tablecolumns{4}
\tablewidth{0pt}
\tablecaption{Parameters of the SPH simulations for Binary Systems with Semi-major Axis of $a_{\rm b}$\label{table:parameters}}
\tablehead{
\colhead{Binary and Disk parameters} & 
\colhead{Symbol} & 
\colhead{Fiducial Value} & 
\colhead{Values} 
}
\startdata
Mass ratio of binary components & $M_{\rm p}/M_{\rm c}$ & 1 & [0.25, 0.5, 1, 2]\\
Binary orbital eccentricity & $e_{\rm b}$ & 0 & [0, 0.2, 0.5]\\
Initial number of particles & $N/10^5$ & 3 & [2, 3, 5, 10]\\
Initial disk mass & $M_{\mathrm{disk}}/(M_{\rm c}+M_{\rm p})$ & 0.001 & 0.001 \\
Initial disk outer radius & $r_{\mathrm{out}}/a_{\rm b}$ & 0.25 & [0.15, 0.25]\\
Initial disk inner radius & $r_{\mathrm{in}}/a_{\rm b}$ & 0.025 & [0.015, 0.02, 0.025]\\
Mass accretion radius & $r_{\mathrm{acc}}/a_{\rm b}$ & 0.025 & [0.015, 0.02, 0.025]\\
Disk viscosity parameter & $\alpha$ & 0.1 & [0.01, 0.1]\\
Disk aspect ratio & $H/r\, (r=r_{\mathrm{in}})$ & 0.035 & [0.02, 0.035, 0.05, 0.065]\\
Initial disk surface density profile $\Sigma \propto r^{-\gamma}$ & $\gamma$ & 1.5 & [0.5, 1.0, 1.5]\\
Initial disk inclination & $i_{\rm 0}$ & $\mathrm{60^{\circ}}$ & [$\mathrm{45^{\circ}}$, $\mathrm{50^{\circ}}$, $\mathrm{55^{\circ}}$,$\mathrm{60^{\circ}}$]\\
\enddata
\label{tab:params}
\end{deluxetable}

Table \ref{tab:params} lists the simulation parameters.  We consider several models whose parameters deviate  from the fiducial model, but not all combinations of parameters were simulated.
Instead, we consider variations of one parameter with all others fixed at the fiducial values, as we did for particle orbits in Figure \ref{fig:testpart}.
For different mass ratios, we also adjust the initial inner and outer disk radii.
We scale the initial disk inner radius from the central binary component by a factor of $(M_{\rm c}/(M_{\rm c}+M_{\rm p}))^{1/3}$. As discussed above, the accretion radius of the central object is taken to be the same as the initial inner disk radius. The inner boundary condition is that any SPH particle that goes inside accretion radius is removed from the simulation. 
The inner boundary radius of the perturbing component is set to $r_{\mathrm{in}} (M_{\rm p}/M_{\rm c})^{1/3}$.
The initial disk outer radius $r_{\mathrm{out}}$ is chosen to be that of a tidally truncated disk in a coplanar binary \citep{Paczynski77}. 
Thus, when we change the mass ratio of the binary, we also change the initial disk outer radius. 
However, in a misaligned binary the tidal torque on the disk is weakened and the disk can expand somewhat beyond this radius \citep{Lubow15}. 
We ignore the effects of disk self-gravity in the simulations presented here. That means the disk does not feel its own gravity, but the stars can feel it.
 We set the ratio of the initial disk mass to the binary total mass to a very low value
of $0.1\%$ such that the binary orbit is hardly affected by disk's gravity.
This mass corresponds to a large initial Toomre Q of $\sim 30$. We will devote a separate paper to studying the effects of higher disk mass and stronger self-gravity.

In post-processing the simulations, we compute the  orbital eccentricity and inclination for each particle using its position and velocity information according to Equations (\ref{eq:e})
and (\ref{eq:i}). We divide the disk into 100 radial bins and calculate the mean properties of the particles within each bin, such as the inclination and eccentricity. The properties of the particles
in these radial bins are not fully independent of each other
because the disk is generally eccentric. Particles residing one bin 
follow streamlines that cover other nearby bins.
Therefore, some correlation of mean properties 
occurs in nearby bins. 
In most of our analysis, we apply fixed radial bins that are fairly widely separated to reduce these correlation effects.
An alternative approach is to apply bins based on particle semi-major axis $a$. We have found that binning in $r$ and $a$ give
similar results for most of our analysis and so we generally report results with bins in $r$
for simplicity.
For the determination of disk warping, however, we adopt bins based on particle semi-major axis, as described later.

The KL period of a test particle (Equation (\ref{eq:klperiod})) depends on the particle's orbital radius. Following Paper I, we determine an analytic estimate of the KL global period of a disk by taking an angular momentum weighted average involving the local KL period of particle orbits
as follows
\begin{equation}
\left < \tau_{\rm KL} \right > \approx \frac{\int_{r_{\rm in}}^{r_{\rm out}}\Sigma r^3 \sqrt{\frac{GM_{\rm c}}{r^3}} dr}{\int_{r_{\rm in}}^{r_{\rm out}}\tau_{\rm KL}^{-1}\Sigma r^3 \sqrt{\frac{GM_{\rm c}}{r^3}} dr},
\label{eq:klperiod2}
\end{equation}
where $\tau_{\rm KL}$ in the denominator is given by Equation (\ref{eq:klperiod}).
This estimate is equivalent to determining the disk global precession frequency as the integrated torque divided by the total disk angular momentum.
This estimate is crude because we apply the angular momentum of circular
orbits, while the disk is generally eccentric. In addition, the analytic particle orbit period
in Equation  (\ref{eq:klperiod}) is itself somewhat approximate, as we found in Section 
 \ref{sec:part}.

\begin{figure}
\centering
\includegraphics[width=0.95\textwidth]{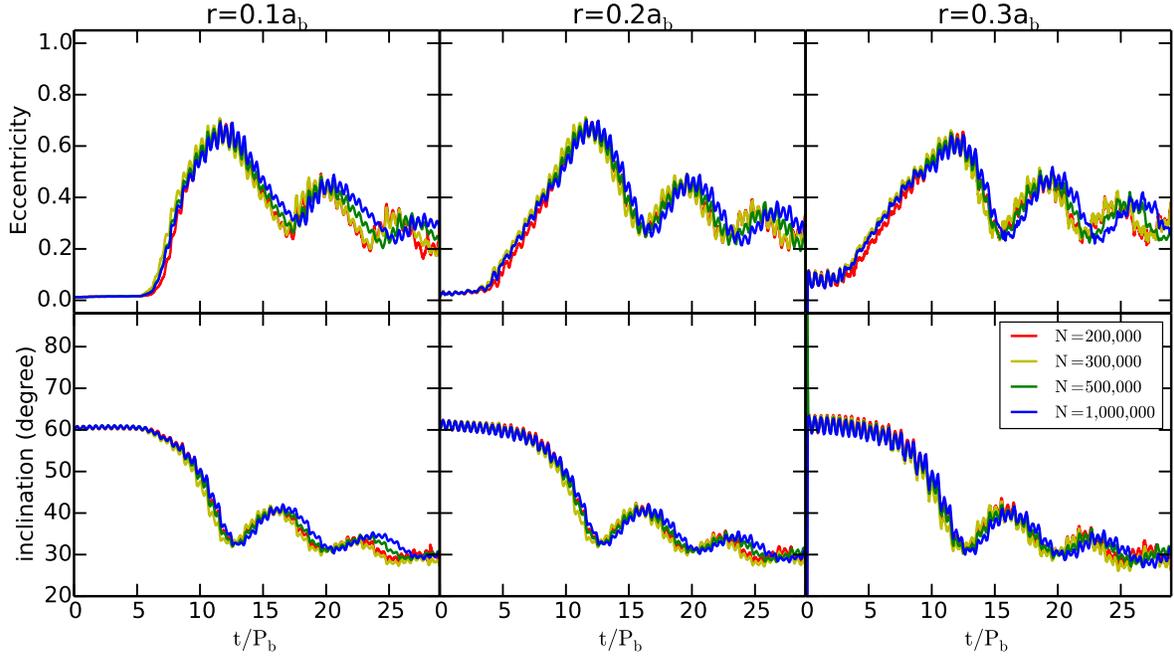}
\caption{Eccentricity (upper row) and inclination (bottom row) evolution of the disk at three different radii from the central star for different initial numbers of particles. $a_{\rm b}$ is
the binary semimajor axis and $P_{\rm b}$ is the binary orbital period. The other parameter values are given in the third column of Table \ref{tab:params}.
(color online) \label{fig:resolution}}
\end{figure}

Figure \ref{fig:resolution} shows the evolution of the disk eccentricity and inclination at three different radii in the disk for various initial numbers of SPH particles. In this figure and following similar figures, disk quantities are taken from three individual radial bins spanning $0.1$, $0.2$ and $0.3 a_b$, respectively. Also, we plot results up to a time of 
$t=28 P_{\rm b}$. At this time, the disk has typically lost $90\%$ of its initial particles. Most of the lost particles are accreted onto the central object. A small fraction of the particles are ejected from the disk, typically less than 10\%. Most of ejected particles are accreted onto the companion and others remain in a circumbinary orbit.
From Figure \ref{fig:resolution}, we see that the effect of the numerical resolution on the disk KL mechanism is in general quite small, especially for the first two cycles. 
Although the cycle period at later times in the low resolution run is slightly shorter than that in high resolution run, all the curves damp to nearly the same level of eccentricity $e \simeq 0.2$ and inclination $i \simeq 28^\circ$  at  a time $t=28 P_{\rm b}$. Given this insensitivity of the overall disk KL cycle pattern to the numerical resolution, we adopt $N=3 \times 10^5$ as the fiducial value
that we use throughout the rest of the paper. 

Figure \ref{fig:resolution} shows some differences from the test particle simulation results in Figure \ref{fig:testpart}. 
We consider how
Equations (\ref{eq:adot}) and (\ref{eq:e-i}) might apply to a disk.
We may expect $a$ and $e$ to represent a measure of the disk's
global semi-major axis and eccentricity, respectively, at some representative radius in the disk, such as the $r=0.2 a_{\rm b}$  that we applied in 
Figure \ref{fig:testpart}.
Locally, a disk can transport angular momentum through waves and viscous
torques. However, globally the vertical (perpendicular to the binary orbit plane) component of the disk angular momentum is conserved in the presence of an axisymmetric ring, apart from losses due to particles leaving the disk.  On the other hand, the disk can dissipate energy, unlike
the case of test particles. Therefore, the evolution of the disk is expected
to depart from that of test particles, as given by Equations (\ref{eq:adot}) and (\ref{eq:e-i}). In Figure \ref{fig:resolution}, we
see that the KL oscillations in the fluid case
undergo damping in which the maxima of eccentricity decrease in time. In addition, 
the eccentricity minima do not drop to zero as in the particle case. The inclination 
drops below the minimum tilt of $39^\circ$ for test particle oscillations. After the KL
oscillations substantially damp, there is a significant residual disk eccentricity $e \sim 0.2$. 

The disk viscous timescale is estimated from Equation
($18'$) of \cite{Lynden-Bell74} 
as $t_v \sim r^2/(12 \nu) \sim 20 P_{\rm b}$
for the fiducial model at $r=0.25 a_{\rm b}$.
We have also verified that the disk density evolution of the fiducial model roughly follows the
results obtained from of a 1D viscous disk model (omits pressure effects)
\cite[e.g.,][]{Lynden-Bell74} for the same initial disk profile and level of viscosity.  Therefore, over the course of the simulations, we
expect the density profile to evolve from the initial one.

The first peak value of the eccentricity seen in the upper row of the plot $e \simeq 0.70$ is approximately consistent with the test particle results 
shown in Figure \ref{fig:testpart} for maximum $e \simeq 0.78$ with an initial tilt of $60^\circ$, although the test particles have a somewhat shorter KL period. 


\begin{figure}
\centering
\includegraphics[width=0.8\textwidth]{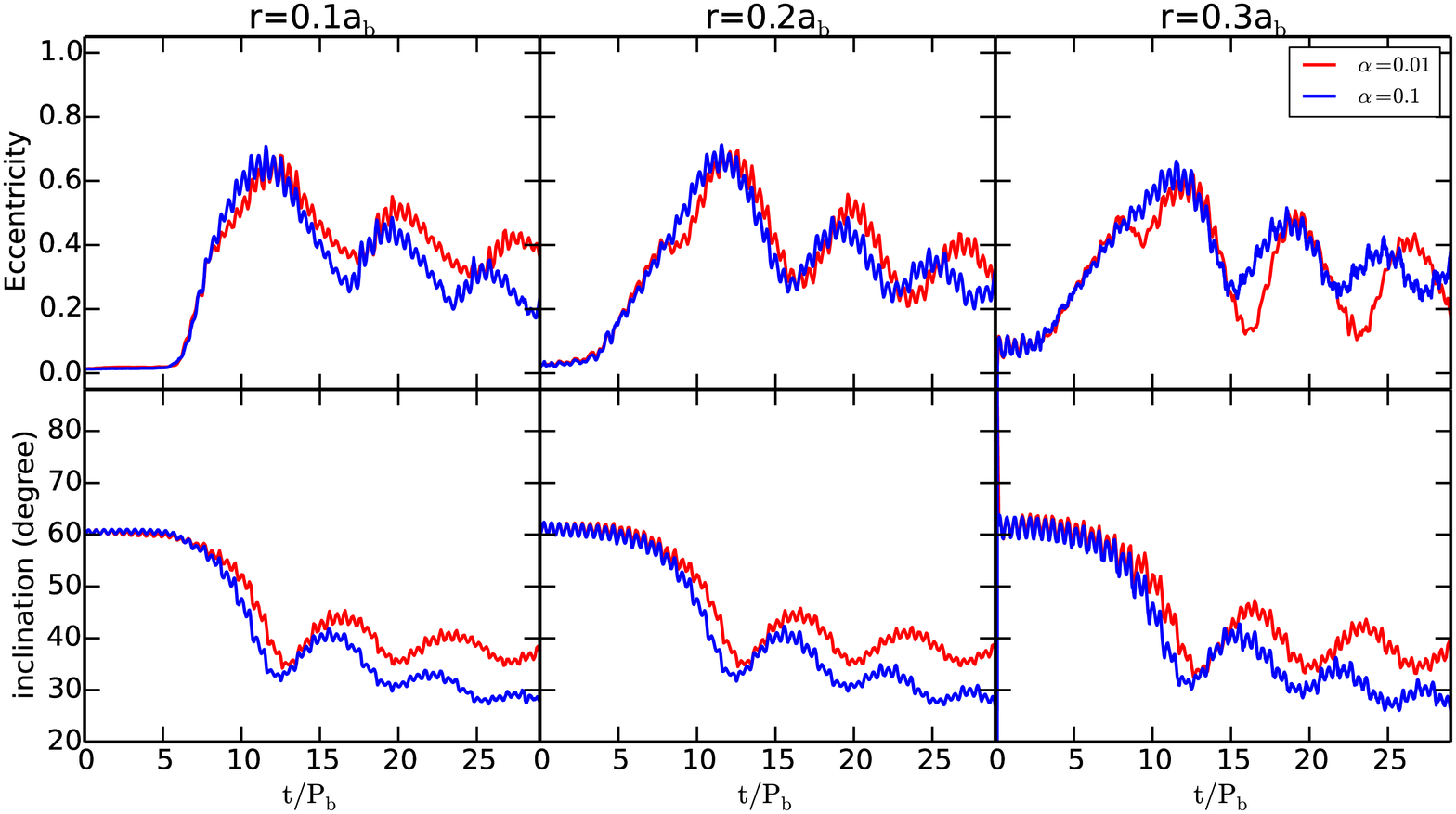}
\includegraphics[width=0.4\textwidth]{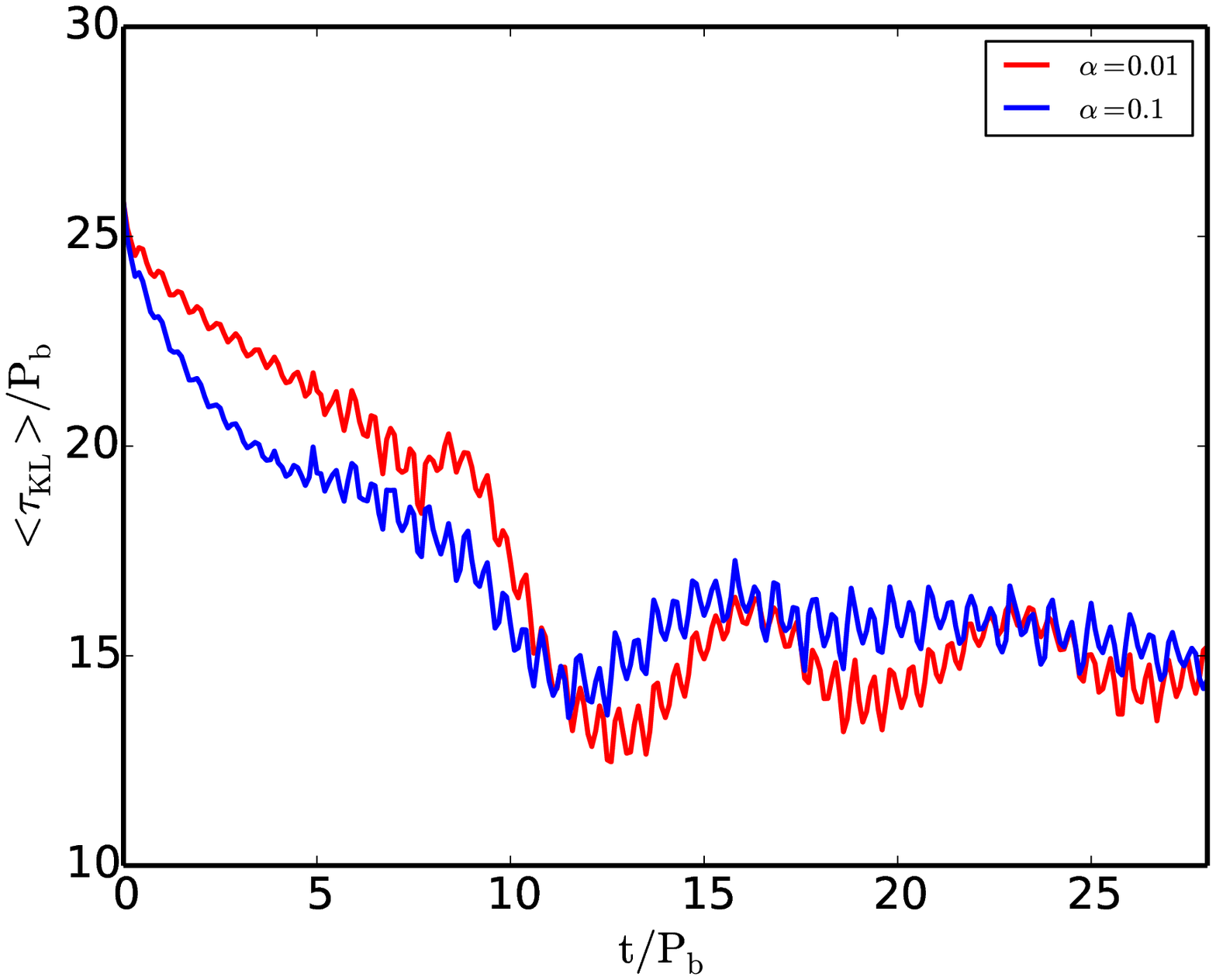}
\caption{The upper figure is similar to Figure \ref{fig:resolution}, but showing the effect of varying the disk viscosity parameter $\alpha$. The lower figure shows the estimated disk KL period as a function of time (according to Equation (\ref{eq:klperiod2})). (color online) \label{fig:viscosity}}
\end{figure}

\begin{figure}
\centering
\includegraphics[width=0.95 \textwidth]{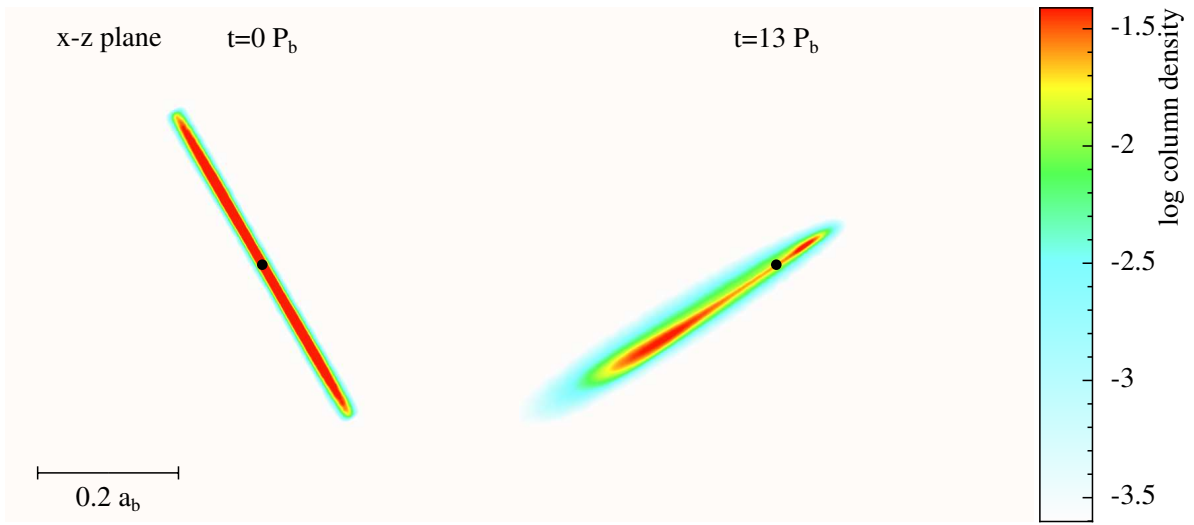}
\caption{View of the fiducial disk toward the $x-z$ plane at times of 0 (left) and 13 $P_{\rm b}$ (right). The binary orbit is in the $x-y$ plane (i.e., the perturbing object moves into and out of the page). 
The central mass is denoted by the black dot. The color coding is for the logarithm base 10 of the column density (i.e., density integrated along the line of sight) in
units of $(M_{\rm c}+M_{\rm p})/a^2_{\rm b}$.
 (Color online) \label{fig:edge_on}}
\end{figure}


\begin{figure}
\centering
\includegraphics[width=0.48\textwidth]{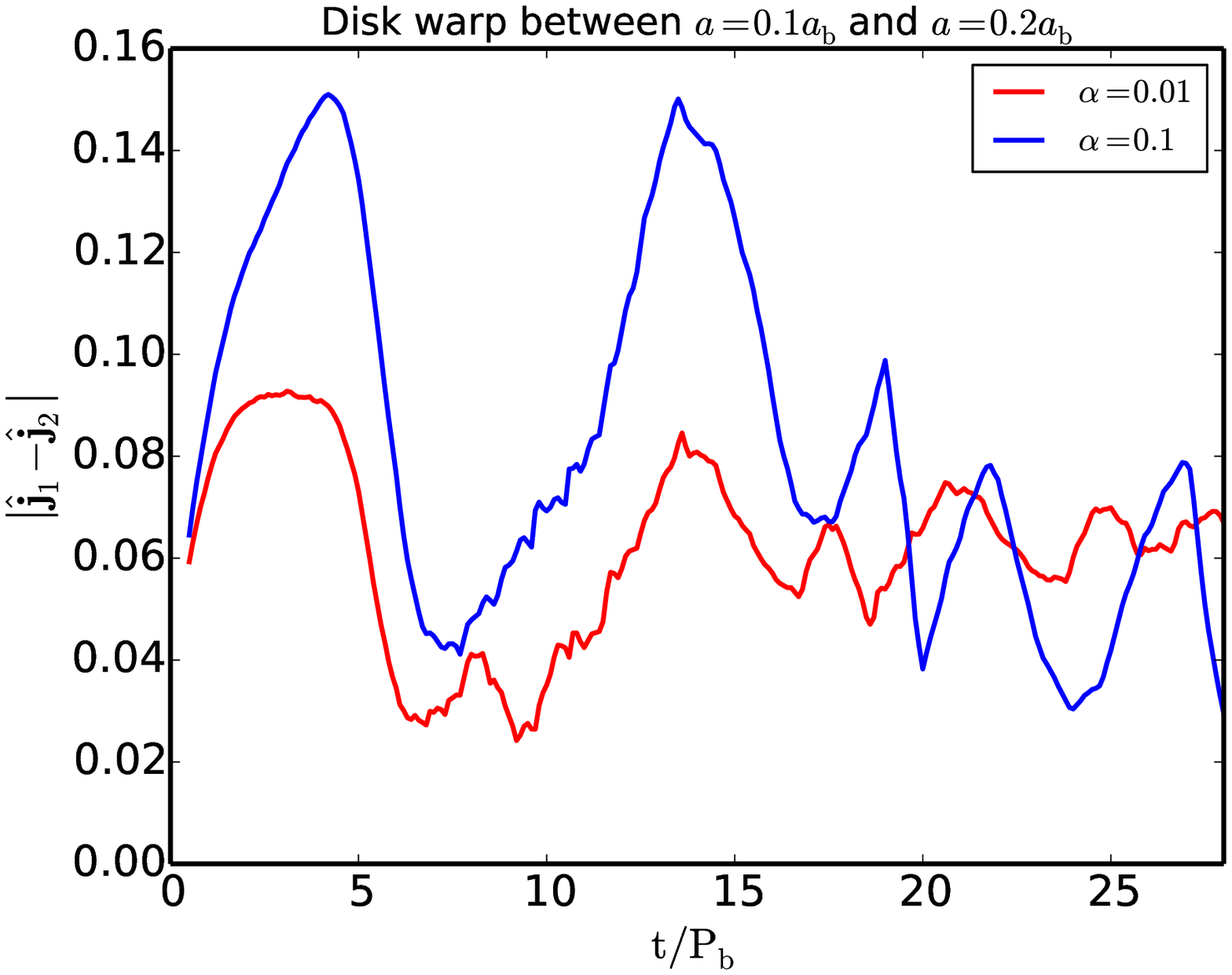}
\includegraphics[width=0.48\textwidth]{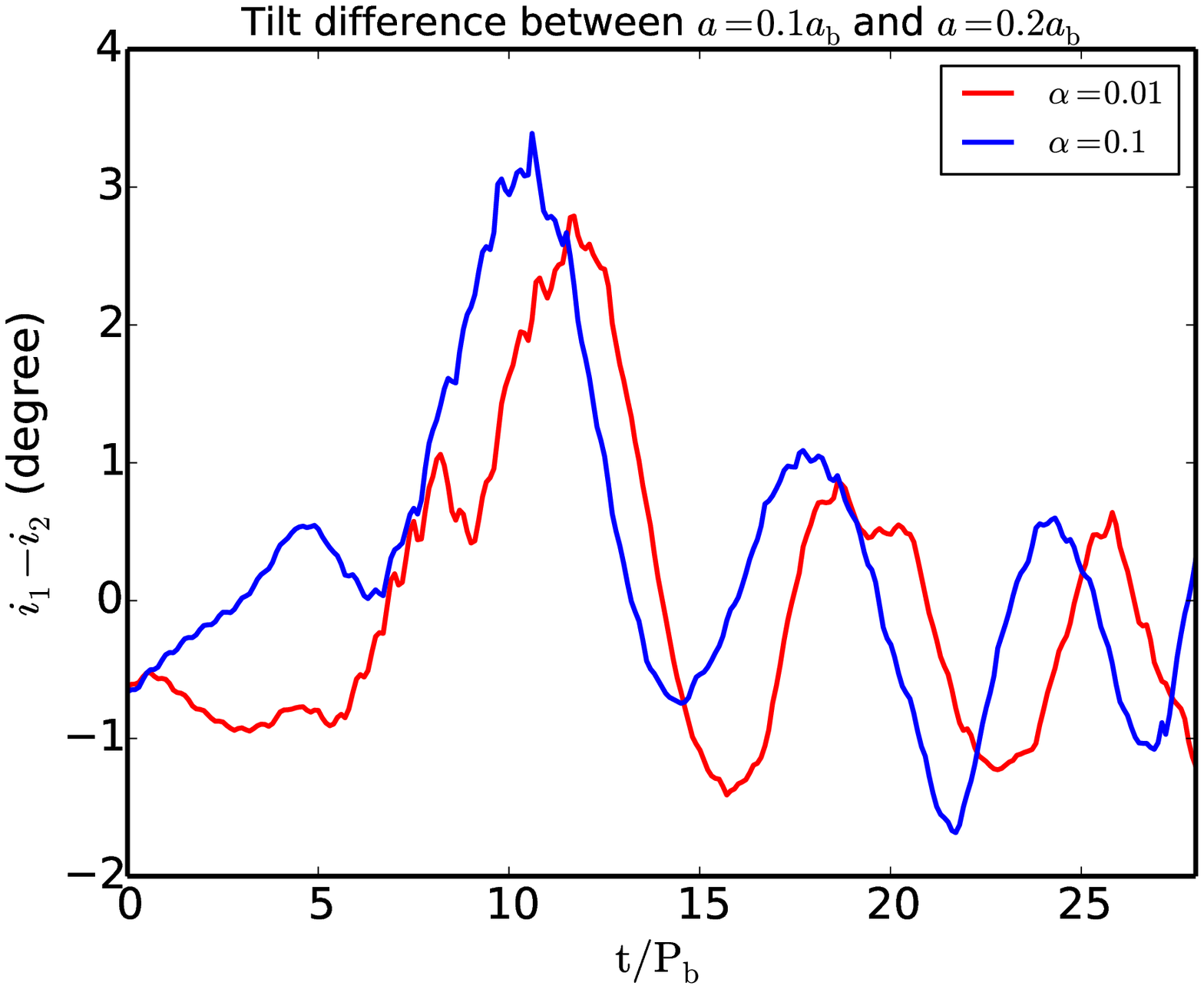}
\caption{Left panel: the magnitude of the difference between the 
average disk angular momentum unit vectors for particles that lie in narrow bins
of semi-major axis values  centered on $a=0.1a_{\rm b}$ and $a=0.2a_{\rm b}$, averaged in time over one binary orbit, 
plotted as a function of time for runs in Figure \ref{fig:viscosity}.  
Right panel: the difference between average disk tilts for particles in narrow bins
of semi-major axis values  centered on $a=0.1a_{\rm b}$ and $a=0.2a_{\rm b}$, averaged in time over one binary orbit,  
plotted as a function of time for runs in Figure \ref{fig:viscosity}.
\label{fig:warp_tilt_a}}
\end{figure}

\begin{figure}
\centering
\includegraphics[width=0.8\textwidth]{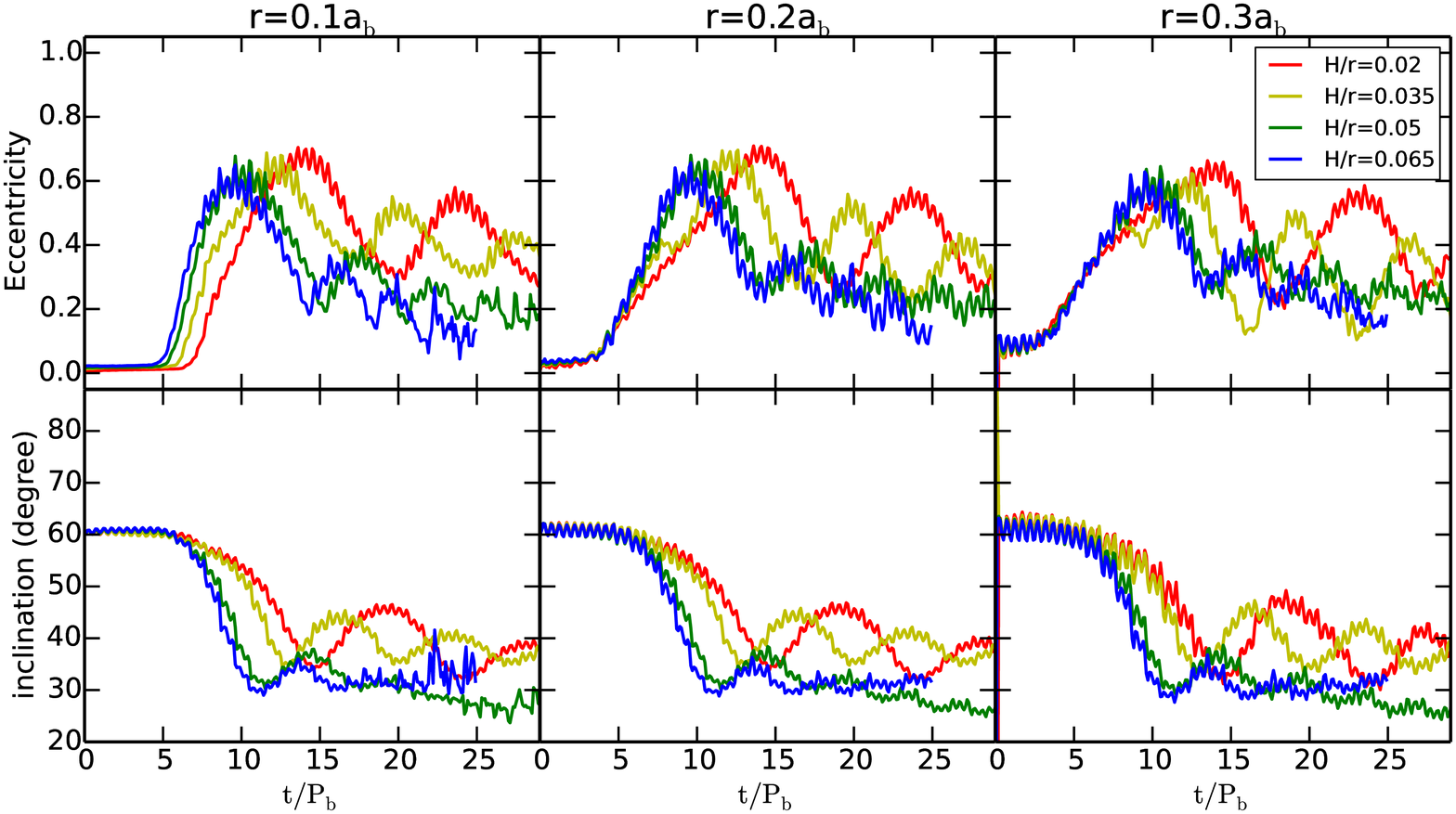}
\includegraphics[width=0.4\textwidth]{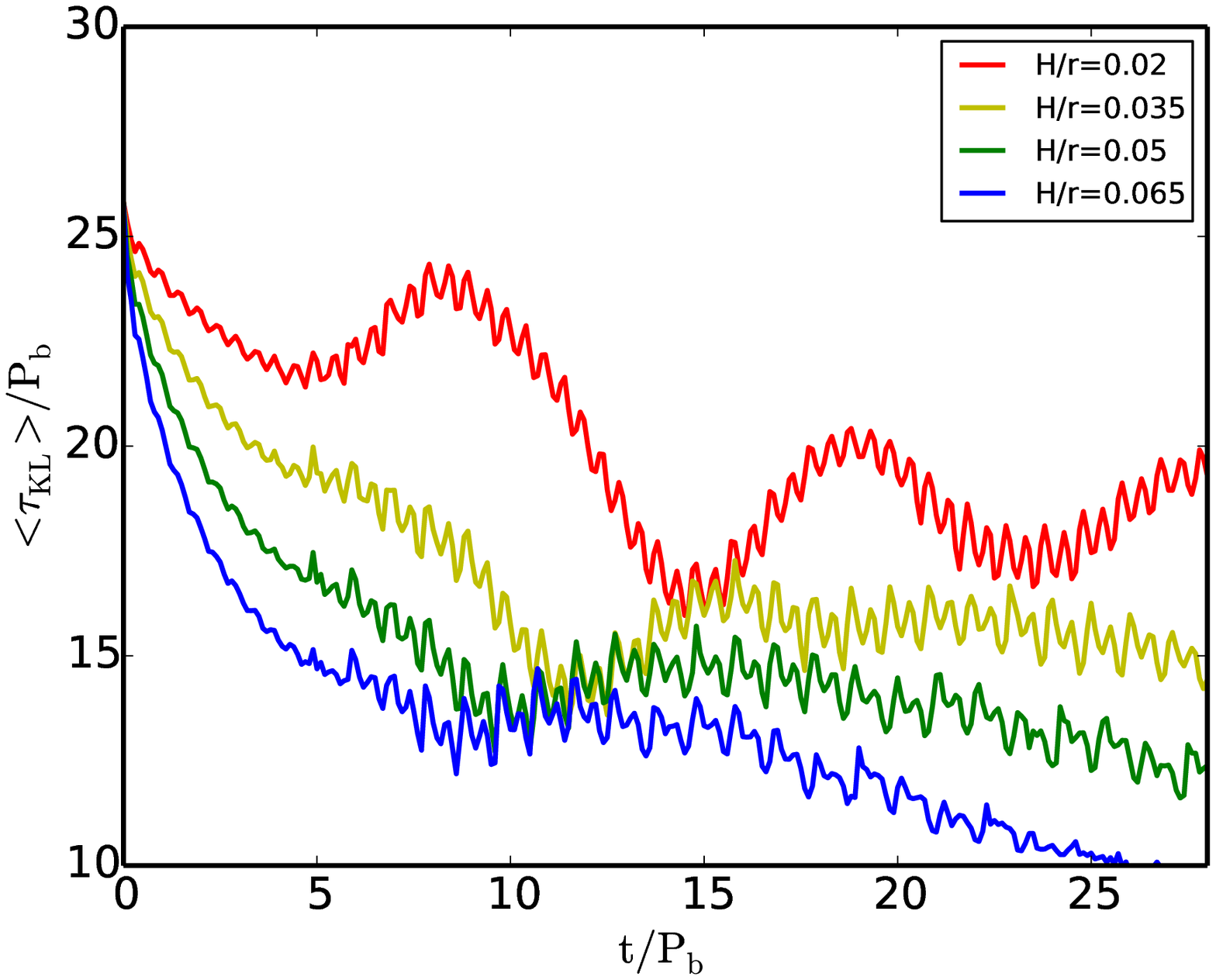}
\caption{Similar to Figure \ref{fig:viscosity}, but showing the effect of varying the initial disk aspect ratio. (color online) \label{fig:hoR}}
\end{figure}

\begin{figure}
\centering
\includegraphics[width=0.8\textwidth]{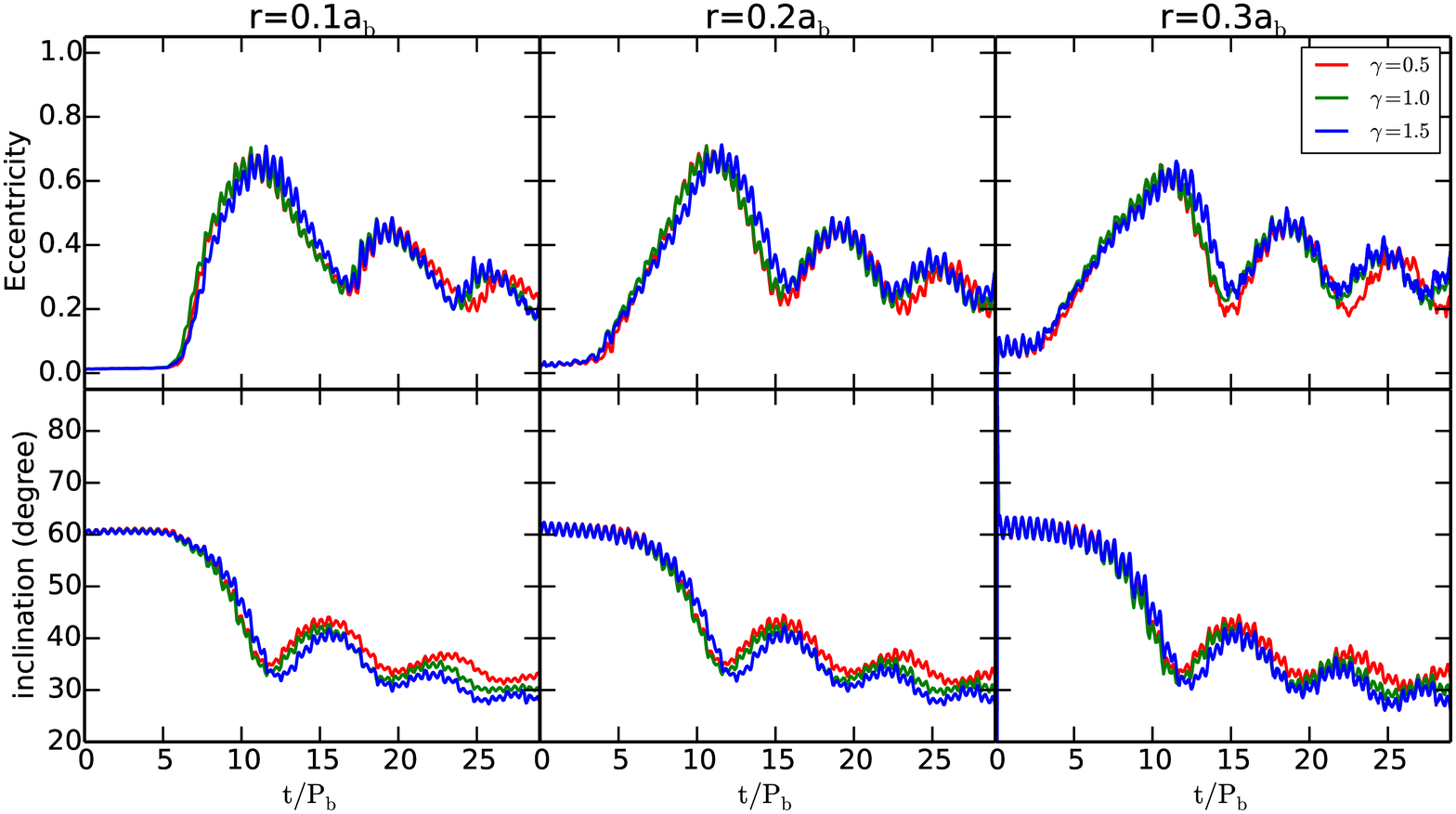}
\includegraphics[width=0.4\textwidth]{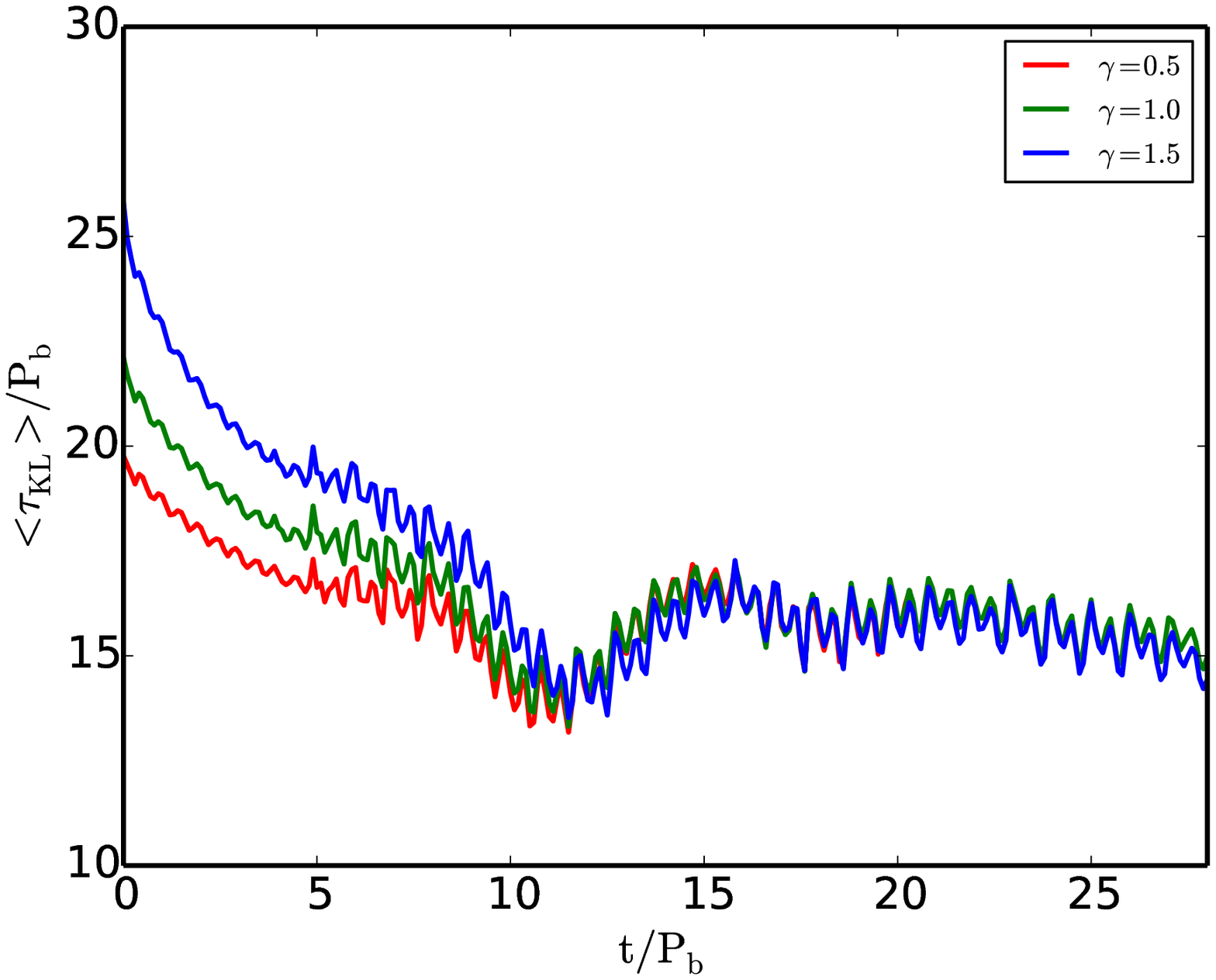}
\includegraphics[width=0.45\textwidth]{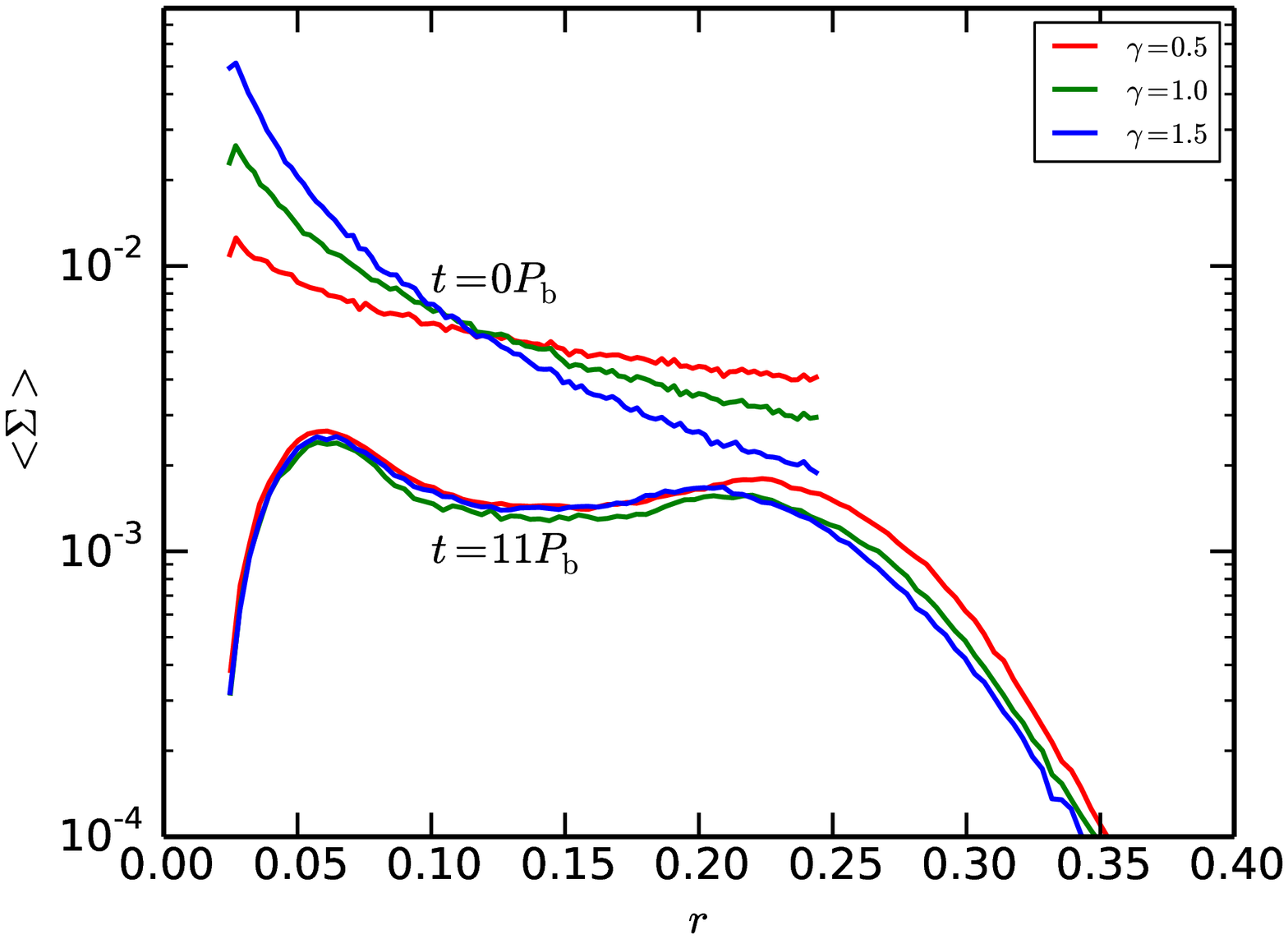}
\caption{Similar to Figure \ref{fig:viscosity}, but showing the effect of varying the initial disk surface density profile. The lower right figure plots the azimuthally averaged surface density profiles at $t=0P_{\rm b}$ and $t=11P_{\rm b}$. 
The surface density is in units of $(M_{\rm c}+M_{\rm p})/a_{\rm b}^2$
and $r$ is in units of $a_{\rm b}$.
(color online)\label{fig:gamma}}
\end{figure}

\begin{figure}
\centering
\includegraphics[width=0.8\textwidth]{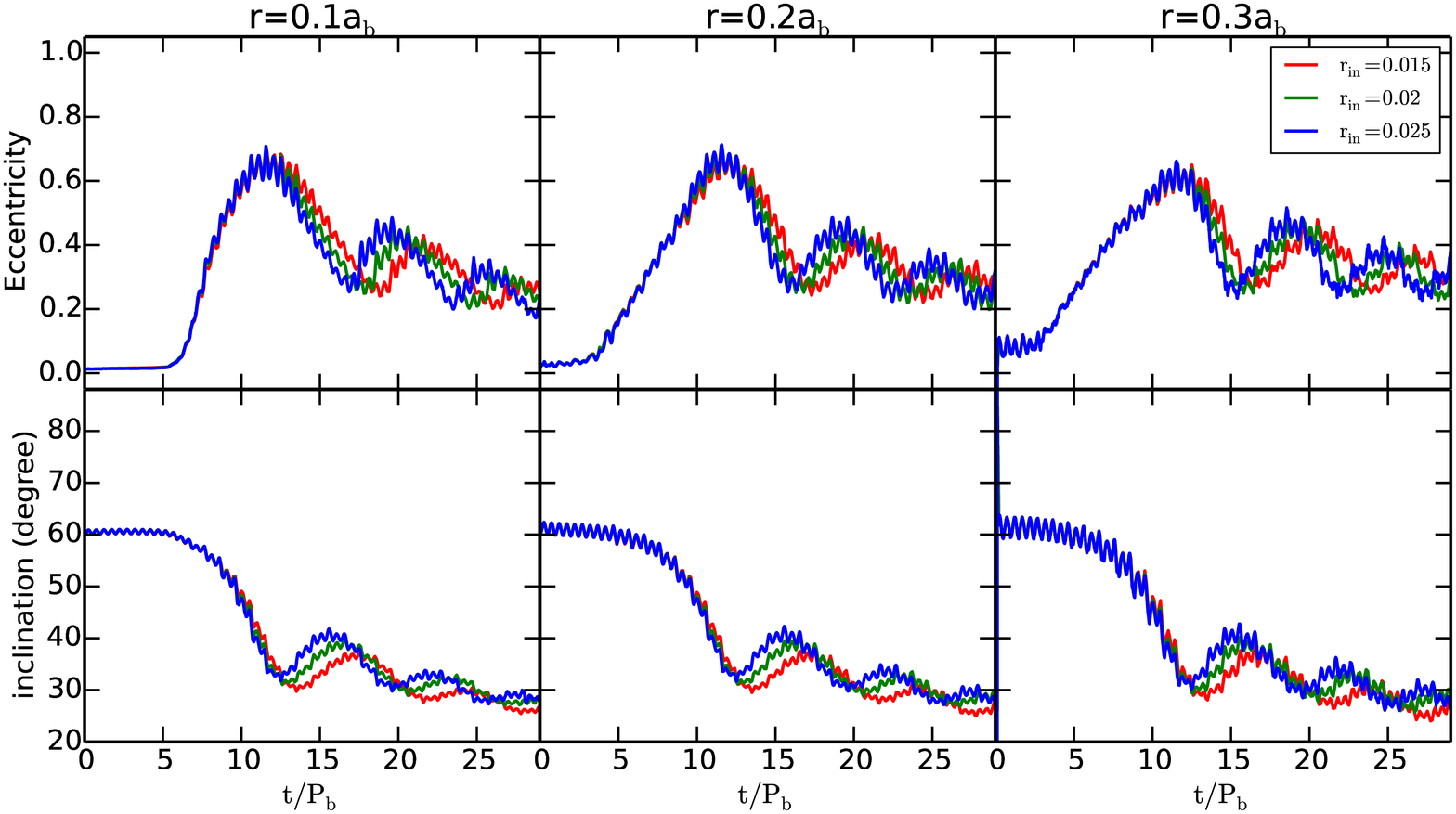}
\includegraphics[width=0.4\textwidth]{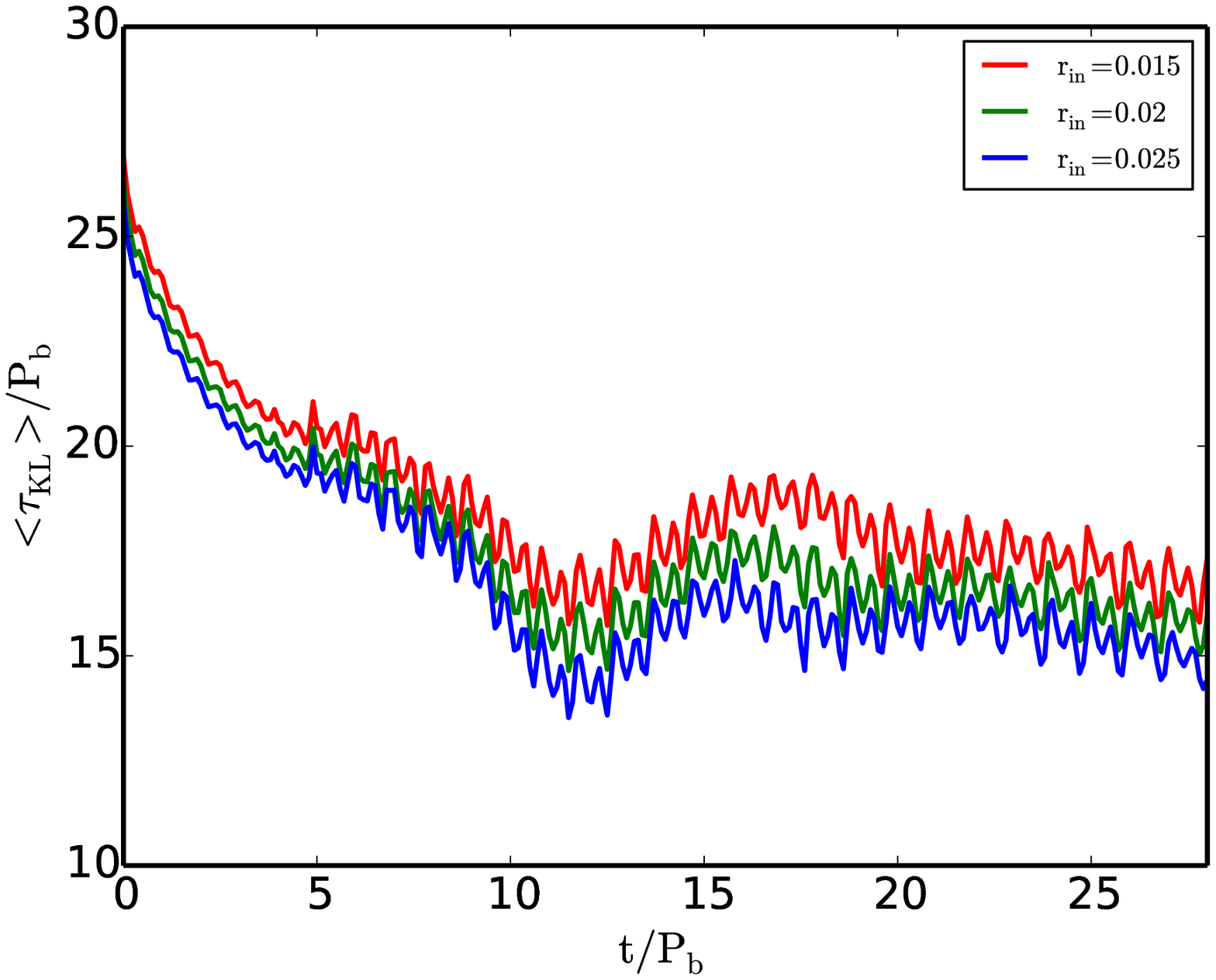}
\caption{Similar to Figure \ref{fig:viscosity}, but showing the effect of varying the initial disk inner radius. (color online)\label{fig:rin}}
\end{figure}

\begin{figure}
\centering
\includegraphics[width=0.8\textwidth]{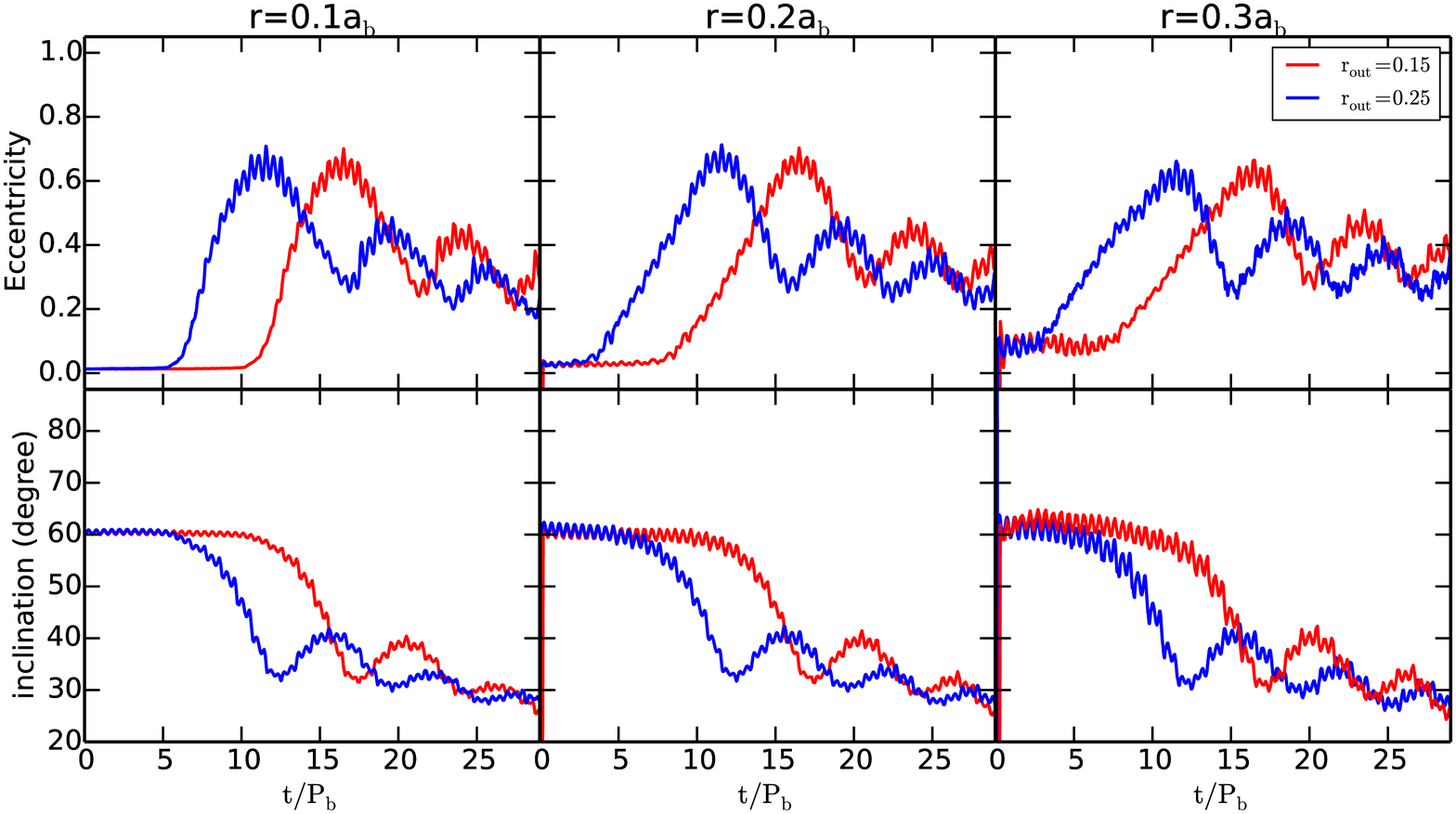}
\includegraphics[width=0.4\textwidth]{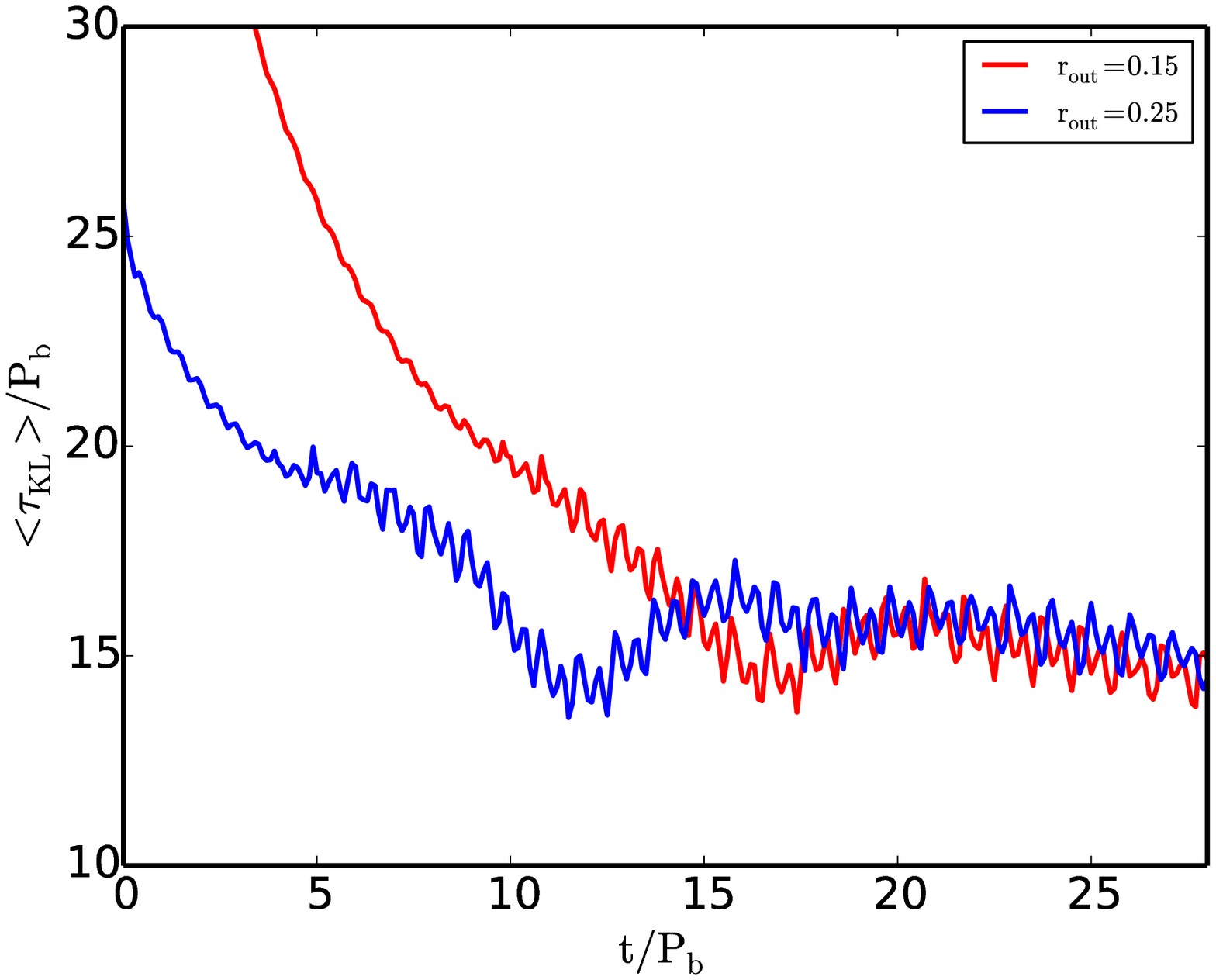}
\caption{Similar to Figure \ref{fig:viscosity}, but showing the effect of varying the initial disk outer radius. (color online) \label{fig:rout}}
\end{figure}

\begin{figure}
\centering
\includegraphics[width=0.8\textwidth]{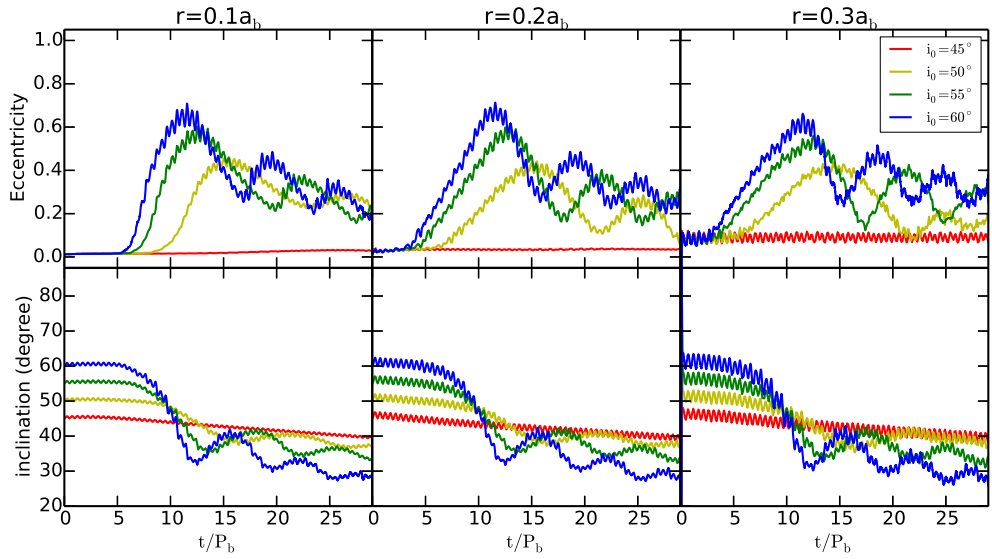}
\caption{Similar to Figure \ref{fig:viscosity}, but showing the effect of varying the initial disk inclination angle. (color online) \label{fig:inc}}
\end{figure}

\begin{figure}
\centering
\includegraphics[width=0.8\textwidth]{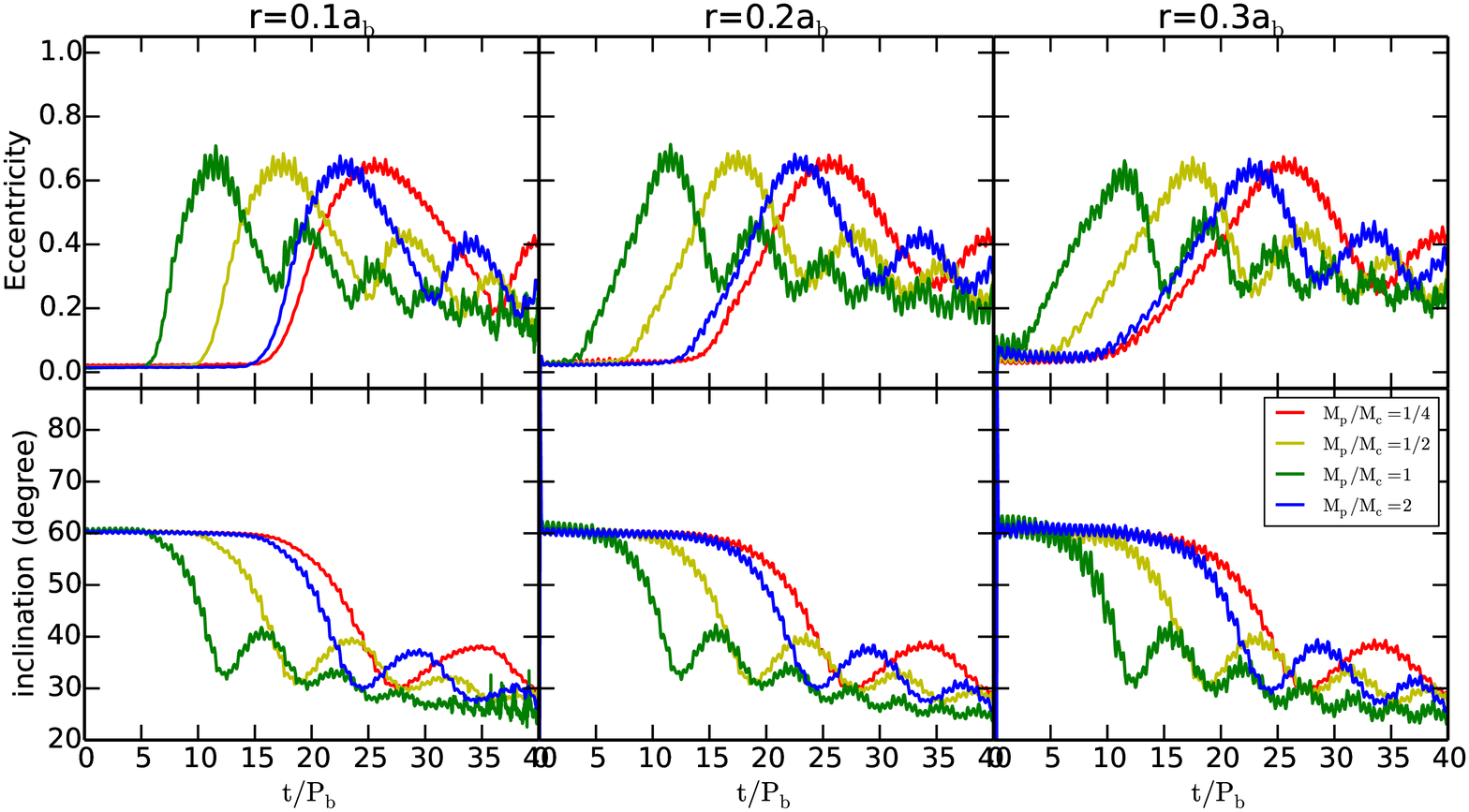}
\includegraphics[width=0.4\textwidth]{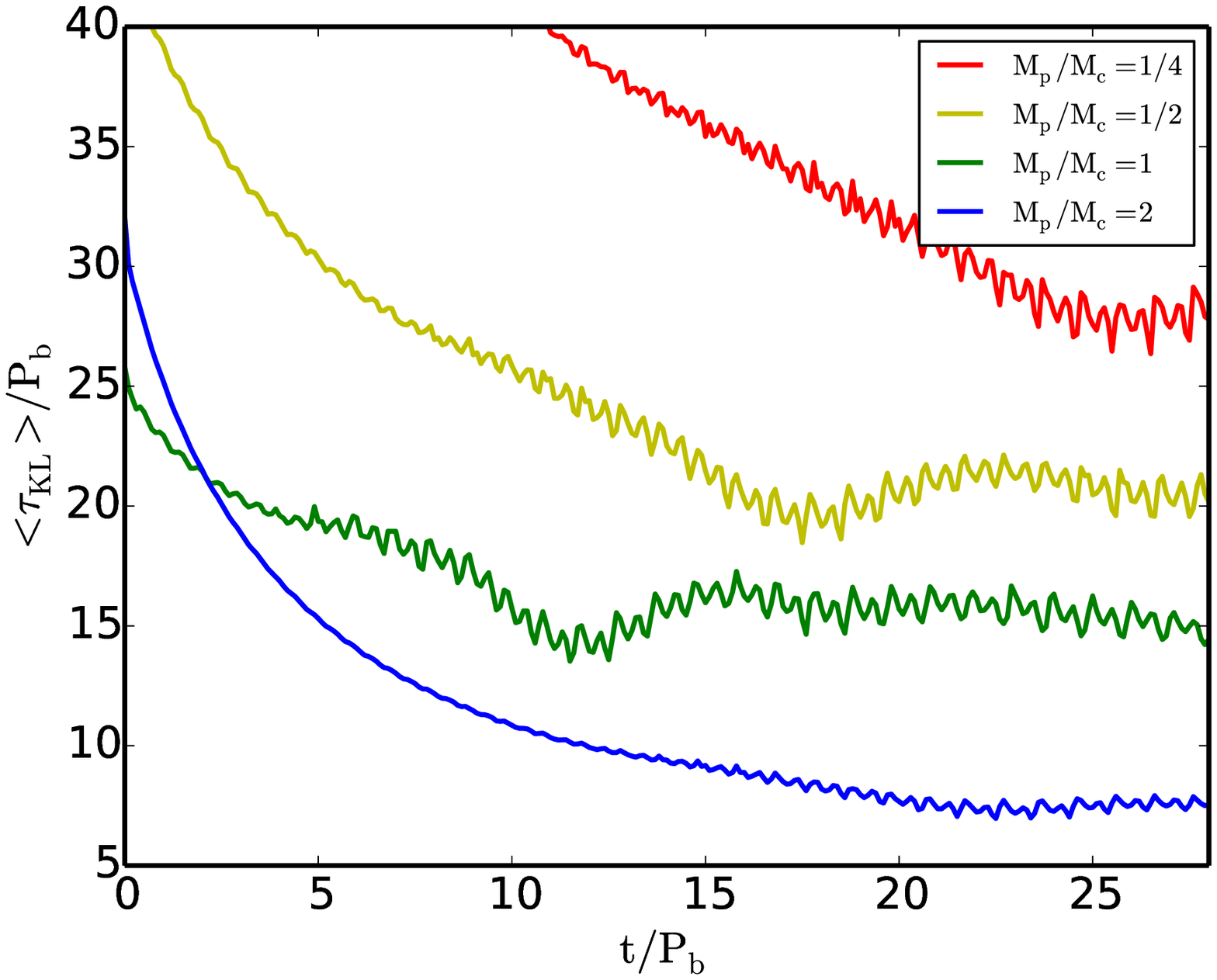}
\caption{Similar to Figure \ref{fig:viscosity}, but showing the effect of varying the binary mass ratio. Note that here we show results up to $t=40P_{\rm b}$ in the upper figure. (color online) \label{fig:mratio}}
\end{figure}

\begin{figure}
\centering
\includegraphics[width=0.95\textwidth]{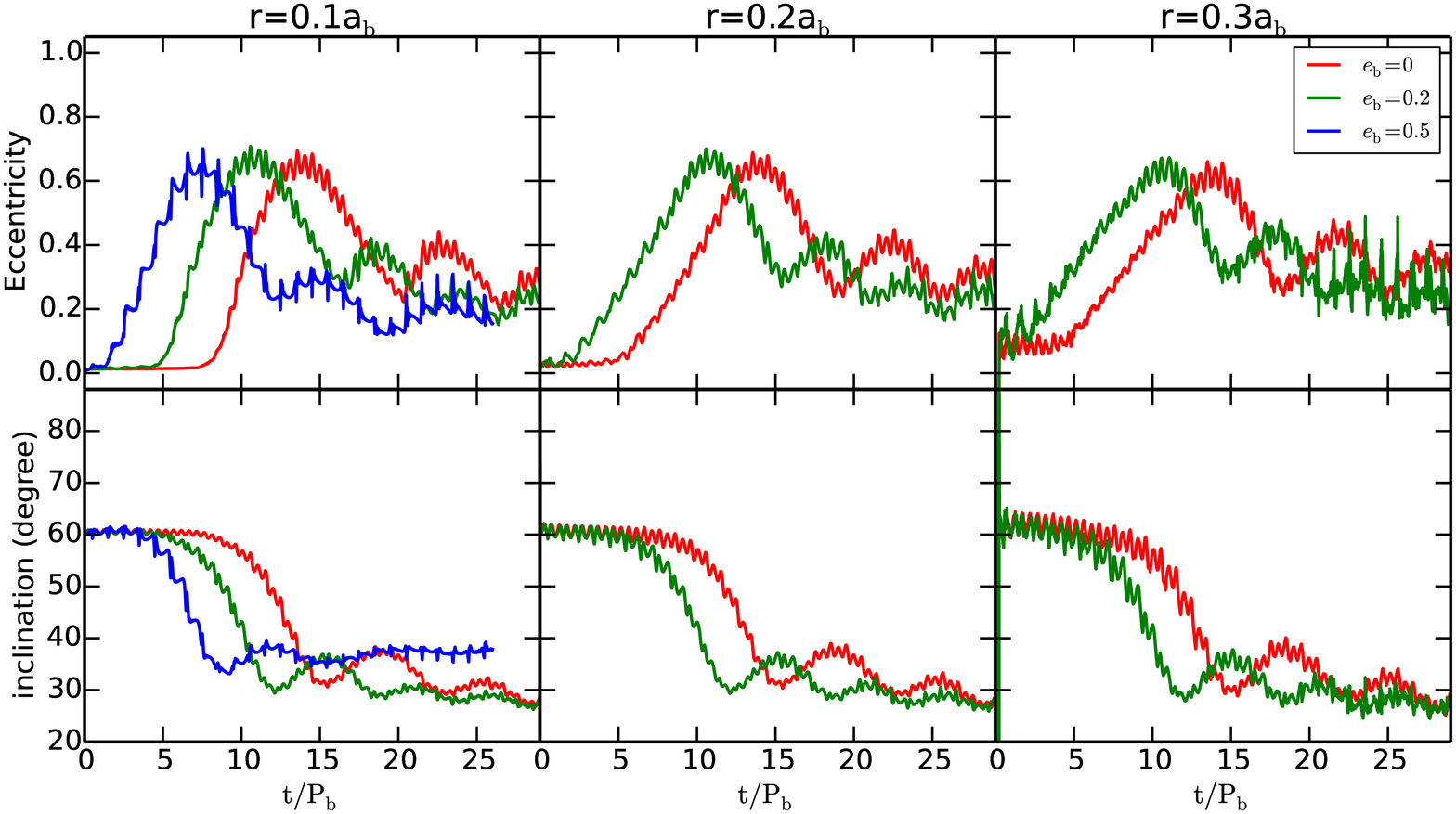}
\caption{Similar to Figure \ref{fig:resolution}, but showing the effect of varying the binary eccentricity. (color online) \label{fig:eb}}
\end{figure}

Figure \ref{fig:viscosity} shows the effects of the disk viscosity on the KL mechanism. For lower viscosity (red curve; $\alpha=0.01$), the time at which the first eccentricity cycle peaks is basically unaffected. 
However, both the eccentricity and the inclination damp at a lower rate because the disk is less dissipative. For $\alpha=0.01$ at later times, the midpoints of the inclination KL oscillations
settle to a tilt that is  near the critical value of $39^\circ$ for KL particle oscillations. Note that there is a substantial remnant disk eccentricity $e \sim 0.3$
after the KL oscillations significantly damp.
The bottom panel of the figure plots the expected KL oscillation periods based on Equation (\ref{eq:klperiod2}). The estimated period is $\sim 15 P_{\rm b}$ that
is to be compared with the period obtained in the SPH simulations (upper panel)
of $\sim 7 P_{\rm b}$. The agreement is crude (factor of 2), likely for reasons
noted below Equation  (\ref{eq:klperiod2}).

The evolution of the disk eccentricity and inclination are well coordinated across different radii of the disk, as is evident from the
similarity of the eccentricity and inclination curves across
different radii in Figures \ref{fig:resolution} and \ref{fig:viscosity}.
Figure \ref{fig:edge_on} shows an edge-on view of the fiducial disk at $t=0$ (left panel) and at $t=13 P_{\rm b}$
(right panel) when
the line of nodes has rotated by about $180^\circ$.
This figure shows that the disk remains fairly flat, but quite lopsided, after substantial nodal and KL phase oscillation changes.

We consider how inclination is radially communicated across the a disk.
A KL disk undergoes tilt oscillations, along with nodal precession.
A disk that is in good radial communication would be expected
to remain nearly flat, that is, have a small degree of warping.
The disk warping is often measured by $r \partial_r \bm{\ell}$
where $\bm{\ell}$ is a unit vector that lies perpendicular to
the local disk orbital plane. In the application to our simulation results, $\bm{\ell}(r)$ would be determined by the average value of $\bm{\ell}$ for particles in a narrow radial bin centered on radius $r$. 
However, as noted above, there are artificial 
correlations in quantities such as $\bm{\ell}(r)$. 
The reason is that eccentric streamlines carry a single value of $\bm{\ell}$ and extend radially in the disk.
A better measure of warping is $a \partial_a \bm{\ell}$ where $\bm{\ell}(a)$
is determined by the average value of $\bm{\ell}$ for particles whose semi-major axes lie in a narrow bin of size $3.5 \times 10^{-3} a_{\rm b}$ centered about semi-major axis  $a$.   The reason for using $a$ is that (noncrossing) streamlines in a fluid disk 
can be uniquely labeled by their $a$ values.
Therefore, $a \partial_a \bm{\ell}$ measures
warping across streamlines.

As an approximate measure of disk warping, 
we compute
\begin{equation}
\delta \hat{j}= |\bf{\hat{j}_2} - \bf{\hat{j}_1}|,
\label{dj}
\end{equation}
where $\bf{\hat{j}_1}$ and $\bf{\hat{j}_2}$ are 
the average of the disk unit angular momentum vectors in bins of particle semi-major axis centered
on $a_1$ and $a_2$, respectively, that are separated by $a_1$ apart.  We choose these values to be $a_1=0.1 a_{\rm b}$ and $a_2=0.2 a_{\rm b}$. 
For the KL effect on test particles in a circular orbit binary,  $a$ is conserved (Equation (\ref{eq:adot})).
By a time $\sim 10 P_{\rm b}$, the value of $\delta \hat{j}$ for test
particles is of order unity, since the outer particle that begins on a circular orbit
with $r_0=a=0.2 a_b$ undergoes a substantial change in tilt, as seen in Figure \ref{fig:testpart}, as well as substantial nodal precession.

Radial communication in a disk is diffusive 
for $\alpha > H/r$ and wave-like for $\alpha < H/r$.
In the diffusive regime, the degree of warping is estimated by equating
the binary
nodal precessional torque to the horizontal (parallel to the disk) component  
of viscous torque  that involves the so-called $\nu_2$ viscosity \citep{Natarajan98}.
The $\nu_2$ effective viscosity owes its existence to the internal flows driven by pressure gradients and limited by (e.g. turbulent) viscosity, as described by \cite{Papaloizou83}. 
We roughly estimate then that the warping due to nodal precession is
\begin{equation}
\delta \hat{j} \simeq \lambda \sin{(i)} \, \alpha \, \left( \frac{\omega_{\rm p}}{\Omega} \right) \, \left(\frac{r}{H} \right)^2,
\label{warp}
\end{equation}
where $\lambda$ is a coefficient of order unity, $\Omega$ is the
circular frequency at the disk radius $r=0.2 a_{\rm b}$, and $\omega_{\rm p}$ is the
nodal precession frequency. 
We have estimated that $\lambda \sim  0.3$ by solving the 1D viscous warp
evolution equation given by Equation (10) of \cite{King13} for the conditions in the fiducial model
that has  $\omega_{\rm p} \sim 0.04 \Omega_{\rm b}$ and find that
$\delta \hat{j} \sim 0.1$, which is roughly similar in value to
the plotted results  in left panel of Figure \ref{fig:warp_tilt_a} for $\alpha =0.1$ (blue curve).

The lower viscosity $\alpha=0.01$ case plotted in left panel Figure \ref{fig:warp_tilt_a}
lies in the wave-like  regime.
The communication is a consequence of radial pressure forces. 
In this simulation
the radial sound crossing timescale
is substantially shorter than the KL oscillation or nodal precessional timescale. 
We have that $r/(c_{\rm s} P_{\rm KL}) \sim 0.1.$ 
As described in Paper 1,  because this ratio is small
the disk undergoes global rigid tilt oscillations \citep{Larwood97, Lubow01}. In this case, the level of warping 
(plotted as the red curve  on the left panel of Figure \ref{fig:warp_tilt_a})
is similar to or weaker than in the viscous regime (blue curve).

 
 The warping $\delta \hat{j}$ depends on the variation of both the disk tilt 
  and direction of the line of nodes. It is affected by both nodal precession
 and KL disk tilt oscillations.
 We consider here the variations of only the disk tilt.
The KL  inclination oscillations can lead to variations in disk tilt across streamlines in
the disk.  
These variations can be expected because the KL oscillation timescale
varies with particle orbit period $P$ and therefore with particle orbit semi-major axis $a$. 
According to Equation (\ref{eq:klperiod}), the KL oscillation period should differ by about a factor of $\sim 3$  in a change from
$a=0.1a_{\rm b}$ to $a=0.2 a_{\rm b}$. Radial communication via viscous and pressure effects
as described above play a role in maintaining a constant tilt.
As seen in the right panel of Figure \ref{fig:warp_tilt_a}, the change in
tilt between $a=0.1 a_{\rm b}$ and $a=0.2 a_{\rm b}$ is small.
The difference in tilt angles between them is typically less than $3^\circ$. This
change is similar to the disk aspect ratio $H/r = 0.035 \simeq 2.0^\circ$, while the disk oscillates over a range of tilt angles
$\simeq 20^\circ$.
In comparing the right panel of Figure \ref{fig:warp_tilt_a} 
with Figure \ref{fig:viscosity}, we see 
that tilt difference is phased with 
the KL cycle.  When the disk inclination is minimum and eccentricity is maximum,
the outer inclination $i_2$ (that has $a = 0.2 a_{\rm b}$) is smaller than the inner inclination $i_1$ (that has $a = 0.1 a_{\rm b}$) .

The effect of varying the disk aspect ratio is shown in Figure \ref{fig:hoR}. Most noticeably, as the disk aspect ratio decreases (from the blue curve to the red curve), the first disk KL cycle gets delayed to a later time. Because the kinematic viscosity $\nu$ varies with $(H/r)^2$, lower values of $H/r$ lead to a slower disk expansion and therefore a weaker perturbation from the companion object.
Lower $\nu$ values also lead to slower damping in both eccentricity and inclination. 
The bottom panel of the figure shows the expected KL oscillation periods based on
Equation (\ref{eq:klperiod2}). As we saw in the case of Figure \ref{fig:viscosity},
the accuracy of this equation is only to about a factor of 2. 
Based on the bottom panel, the ratio of KL periods for $H/r=0.065$ to $H/r=0.02$
is about 1.3, while the ratio in the SPH simulations (upper panel) is about 1.5.

We present in Figure \ref{fig:gamma} three runs using different power-law indices for the initial disk surface density profile. The differences in the results for these three cases is found to be quite small.  The bottom left panel of this figure
shows that the expected KL periods based on Equation
(\ref{eq:klperiod2}) are about the same after a time of 
about $10 P_{\rm b}$, as is in agreement with the SPH simulations in the upper panels. The bottom right panel shows that after initial adjustment, the
density distributions are very similar.

In Figure \ref{fig:rin}, we show the effects of changing the disk inner radius. 
The first disk KL cycle does not seem to be affected by the location of the disk inner boundary. However, the cycle period does get slightly longer as the disk inner radius decreases. 
The bottom panel of this figure
shows that the expected KL periods based on Equation
(\ref{eq:klperiod2}) are in qualitative agreement with this trend. 
This effect is likely be the result of having more material located further from the companion, closer to the central object, resulting
in weaker global perturbations. 

Figure \ref{fig:rout} shows the effects of changing the disk outer radius. 
With smaller initial disk outer radius, the disk initially responds more slowly since the perturbations from the companion object are weaker. 
Thus, the effect in this case is simply a time delay in the cycles.
The bottom panel of this figure
shows that the expected KL periods based on Equation
(\ref{eq:klperiod2}) are in agreement with this description. 
The case with the smaller initial radius has an initial KL period
that is much longer than for disk of large initial radius. By a time of $\sim 15 P_{\rm b}$,
both cases have the nearly same KL periods, as seen in the SPH simulations results
of the upper panels. 


Figure \ref{fig:inc} shows the effects of changing the initial disk inclination. As $i_0$ varies from $60^{\circ}$ to $45^{\circ}$, the KL cycles occur at a progressively later time with a longer cycle period and smaller oscillation magnitude. This trend seems to be in line with the test particle results
discussed in the previous section, where the KL oscillation period increased with decreasing inclination. Equation (\ref{eq:maxe}) predicts eccentricity maxima of $0.76$, $0.67$, and $0.56$ for initial tilts of $60^\circ$, $55^{\circ}$, and $50^{\circ}$, respectively, while Figure \ref{fig:inc} exhibits maximum $e$ values during the first KL cycle of $0.7$, $0.6$, and $0.4$. In the test particle case, for an initial tilt of $i_0=45^{\circ}$, the
eccentricity can reach values of $0.4$, based on Equation (\ref{eq:maxe}), or 0.3, based on simulations (top-left panel of Figure \ref{fig:testpart}). 
However, for this initial inclination angle, the KL mechanism barely operates in the disk whereas it is still relatively strong in the test particle runs (see the left column of Figure \ref{fig:testpart}). Thus, the critical initial tilt angle for the onset of KL oscillations in a hydrodynamical disk is somewhat higher than for a test particle ($39^\circ$). On the other hand, in the first disk KL cycle, the inclination can drop  below $39^{\circ}$, which does not occur in the particle case.  

Figure \ref{fig:mratio} demonstrates the effects of varying the binary mass ratio. For $M_{\rm p}/M_{\rm c} \leq1$,  a smaller mass ratio leads to a later occurrence of the KL cycle and also a longer cycle period (from the green curve to the red curve), which again is in line with the test particle results. 
The bottom panel of this figure
shows that the expected KL periods based on Equation
(\ref{eq:klperiod2})  at a time of $25 P_{\rm b}$ are about 
$30 P_{\rm b}$, $20 P_{\rm b}$, $15 P_{\rm b}$, and $7 P_{\rm b}$ for mass ratios of $1/4$, $1/2$, $1$, and 2, respectively, whereas the SPH simulation results show 
$15 P_{\rm b}$, $11 P_{\rm b}$, $8 P_{\rm b}$, and $11 P_{\rm b}$. 
For both approaches for mass ratios less than or equal to unity, the KL period decreases with mass ratio, as expected. The period for the corresponding
cases are again different by about a factor or 2. 
In the case where
the perturbing companion is more massive with a mass ratio of 2, the period in the SPH simulation
does not follow this trend and instead increases.

 The effects of varying the binary eccentricity are shown in Figure \ref{fig:eb}.
 The initial disk outer radius is $r_{\rm out} = 0.2 a_{\rm b}$ instead of the fiducial value of $0.25 a_{\rm b}$. 
 This smaller initial  disk outer radius was used because the increased binary eccentricity
 causes a stronger response in the disk and a smaller disk truncation radius, as is known to occur for the coplanar case
 \citep{Artymowicz94}.
 Other parameters are set to their fiducial values.
 We do not plot the results for $e_{\rm b} = 0.5$ at $r=0.2 a_{\rm b}$ and $r=0.3 a_{\rm b}$
 because the disk is truncated within this radius due to the smaller periastron
 distance. 
 The KL oscillation periods are about the same in the three cases, within about 10\% of each other.
 The maximum eccentricities are also about the same in the three cases, as we found for test particles
 in Section \ref{sec:part}. In the $e_{\rm b}=0.5$ case, the inclination decays to a value close to $i_{\rm cr}=39^\circ$,
 while the lower binary eccentricity cases decay to a smaller tilt angle.


\section{ Discussion and Summary \label{sec:sum}}

In this paper, we investigated the conditions under which KL oscillations
can operate on a fluid disk, as was first described in \cite{Martin14b}.
Such a disk undergoes a complicated evolution involving tilt and eccentricity
oscillations, as well as the usual nodal and apsidal precession.
We carried out this analysis
by means of SPH simulations that include the effects of disk pressure and viscosity. We considered a range of disk viscosities (Figure \ref{fig:viscosity}),
sound speeds (Figure \ref{fig:hoR}), initial density profiles (Figures \ref{fig:gamma}, \ref{fig:rin}, \ref{fig:rout}), initial inclinations 
(Figure \ref{fig:inc}), binary mass ratios (Figure \ref{fig:mratio}),
and binary eccentricities (Figure \ref{fig:eb}).
We found that the KL effect generally operates over a range of
of disk and binary properties. 

The general picture we have is that an initially circular circumstellar disk that is sufficiently inclined ($\sim 45^{\circ}$ -- $135^{\circ}$) with respect to the orbit plane of a binary undergoes damped KL eccentricity and tilt oscillations. During these oscillations, the disk can develop substantial eccentricities, $e \ga 0.5$. Once the KL oscillations have mostly damped, the disk achieves a tilt in the prograde disk case  (initial tilt less than $90^{\circ}$)   $i \simeq 39^{\circ}$ or somewhat smaller and an eccentricity of up to a few tenths. Over longer timescales, if the disk survives that long, we expect that the disk eccentricity will eventually damp \citep[e.g.,][]{Ogilvie01a} and the disk will become coplanar with the binary orbit plane \citep[e.g.,][]{King13} due to viscous effects.

The properties of KL disk oscillations are somewhat similar to
the properties of KL test particle oscillations (Figure \ref{fig:testpart}).
In both cases, inclination and eccentricity are interchanged.
The peak eccentricity achieved by a disk in the first KL oscillation
is generally close to the eccentricity achieved by test particles for the same initial inclination.
In addition, for smaller mass perturbers, the KL oscillations are slower for
both test particles and disks
(Figure \ref{fig:mratio}).

However, there are some significant differences between a fluid disk and test particles.
The KL oscillation frequency of test particles varies with particle orbit semi-major axis
as $a^{3/2}$.
On the other hand, a disk undergoes a large scale global oscillation with a small
level of warping for typical parameters (Figures \ref{fig:edge_on} and \ref{fig:warp_tilt_a}). 
For a circular
orbit binary, the inclination of prograde test particle orbits relative to the binary orbit plane  
must be greater than the critical angle $i_{\rm cr}=\arccos{(\sqrt{3/5})} \simeq 39^\circ$
for KL oscillations to occur. In the case of a disk, we find that the minimum
inclination angle appears to be somewhat larger ($\ga 45^{\circ}$)  (see Figure  
\ref{fig:inc}). 
For a circular
orbit binary, the inclination of prograde test particle orbits oscillates
between the initial value $i_0 > i_{\rm cr}$ and $i_{\rm cr}$.
For the case of a disk, however, the oscillation can dip below $i_{\rm cr}$
(e.g., Figure \ref{fig:viscosity}). Therefore, even though a disk
requires a larger tilt to operate than a test particle orbit,
the disk oscillations can reach lower tilt angles.

Another major difference between test particles and disks is that
disks can dissipate energy. As a result, the KL oscillations damp,
but leave a residual disk eccentricity that we find to be typically a few tenths. 
We find that the damping rate
depends on the level of disk viscosity (Figure \ref{fig:viscosity}).
The disk eccentricities may also damp through a decay process involving
a parametric instability \citep{Papaloizou05a, Papaloizou05b, Barker14, Ogilvie14}.
Our simulations may not be capable of resolving this instability that
occurs on scales smaller than the disk thickness.
It is also possible that the initial KL oscillation damping involves shocks,
since the streamlines may attempt to cross at high eccentricity.
We have found indirect evidence for such effects in some simulations
that we carried out with a low value of viscosity parameter $\beta_{\rm AV}$.
As mentioned in Section \ref{sec:kld},
this parameter is used in the SPH code to prevent interparticle penetration that
artificially occurs in the code where shocks are present. For small values of $\beta_{\rm AV}$,
the SPH simulations displayed unphysical behavior, symptomatic
of interparticle penetration.
Over longer timescales in the post-KL oscillation phase, the disk is expected to eventually align
with the binary orbit plane through viscous effects \cite[e.g.,][]{King13}.

For an eccentric orbit binary, the KL test particle oscillations can
reach to higher inclination angles than the initial tilt. Inclination angles can
reach $90^{\circ}$ and even higher, resulting flipping from prograde to retrograde
orbits \citep{Lithwick11, Naoz11, Naoz13a, Naoz13b, Liu15}. In reaching $90^{\circ}$, the orbit becomes radial.
KL disk oscillations cannot achieve a similar increased
level of inclination (certainly not to eccentricity close to unity) because of dissipation. 

In this paper, we have omitted the effects of disk self-gravity.
Such effects may suppress KL oscillations for sufficiently massive disks
\citep{Batygin12, Martin14b}. We intend to describe the influence of self-gravity
in a future paper.

Confirmation of the KL mechanism acting on a disk would come from an observation of an eccentric and misaligned disk in a binary system. The KL oscillations are applicable to binary systems on all scales. For example, the oscillations could play a role in planet formation and evolution around one component of a binary star.   There is observational evidence that some exoplanets may be undergoing KL cycles in wide binaries \citep[e.g.][]{Wu03,Takeda05}. In future work, we plan to investigate the evolution of a planet--disk system that orbits around one component of a binary system in order to understand how these planets can form. 

Be/X-ray binaries may also be influenced by the KL oscillations of a misaligned disk \citep{Martin14a,Martin14b}. Be stars may form disks from material that is ejected from the equator of the rapidly rotating star \citep{Lee91,Hanuschik96,Carciofi11}. Be/X-ray binaries contain a Be star with a rotation axis that is misaligned to that of the orbit of a companion neutron star \citep[see Table 1 in][for some observations]{Martin11}. We found in our simulations that roughly 10\% of the initial disk mass
is captured by the companion object for the fiducial model. As a disk undergoes KL oscillations to maximum eccentricity,
its apastron radius grows  which allows this capture to occur. 
This mass exchange may explain the so-called giant Type II
outbursts as being due the accretion of disk mass by the companion neutron star \citep{Martin14a}.


Finally, the KL oscillations of a disk may also be relevant to SMBH binaries. It is possible that the circumbinary disk around a SMBH binary will be misaligned \citep[e.g.][]{Nixon11a,Nixon11b}. Thus, as gas accretes inwards from the circumbinary disk, misaligned disks will form around each SMBH. These disks will be unstable to KL oscillations if the misalignment angle is sufficiently large. This effect could have implications for the subsequent disk evolution and star formation in these systems.


W.F. and S.H.L. acknowledge support from NASA grant NNX11AK61G.  Computing resource supporting this work was provided by the institutional computing program at Los Alamos National Laboratory. We thank Daniel Price for providing the \texttt{PHANTOM} code for SPH simulations and \texttt{SPLASH} code \citep{Price07} for data analysis and rendering of figures.


\end{document}